\documentclass[twocolumn,superscriptaddress,aps,longbibliography,nofootinbib]{revtex4-2}

\usepackage[utf8]{inputenc}
\usepackage{amsmath,amsthm,amssymb,amsfonts}
\usepackage{braket}
\usepackage{graphicx}
\usepackage{multirow}
\usepackage[breaklinks=true,colorlinks=true,linkcolor=blue,urlcolor=blue,citecolor=blue,pagebackref=false]{hyperref}
\usepackage{color}
\usepackage{times}
\usepackage{float}
\usepackage{multirow}
\usepackage{bbold}
\usepackage[abs]{overpic} 

\usepackage[normalem]{ulem} 

\usepackage{ulem}

\usepackage{tikz}
\usetikzlibrary{quantikz}

\date{\today}

\usepackage{natbib}
\usepackage{graphicx}

\usepackage{tabularx}

\newcounter{protocol}
\newenvironment{protocol}[1]
  {\par\addvspace{\topsep}
   \noindent
   \tabularx{\linewidth}{@{} X @{}}
    \rule{0.485\textwidth}{.4pt}
    \refstepcounter{protocol}\textbf{Protocol \theprotocol} #1 \\
    \rule{0.485\textwidth}{.4pt}}
  { \\
   \rule{0.485\textwidth}{.4pt}
   \endtabularx
   \par\addvspace{\topsep}}

\newcommand{\sbline}{\\[.5\normalbaselineskip]}

\newcommand{\smallsim}{\smallsym{\mathrel}{\sim}}
\newcommand{\smallbrl}{\smallsym{\mathrel}{(}}
\newcommand{\smallbrr}{\smallsym{\mathrel}{)}}

\newcommand{\ul}{U_\mathrm{L}}
\newcommand{\ur}{U_\mathrm{R}}
\newcommand{\uld}{U_\mathrm{L}^\dagger}

\newcommand{\mycomment}[1]{}

\makeatletter
\newcommand{\smallsym}[2]{#1{\mathpalette\make@small@sym{#2}}}
\newcommand{\make@small@sym}[2]{%
  \vcenter{\hbox{$\m@th\downgrade@style#1#2$}}%
}
\newcommand{\downgrade@style}[1]{%
  \ifx#1\displaystyle\scriptstyle\else
    \ifx#1\textstyle\scriptstyle\else
      \scriptscriptstyle
  \fi\fi
}
\makeatother

\begin{document}

\title{Cross-Platform Verification in Quantum Networks}

\author{J.\ Knörzer}
\affiliation{Institute for Theoretical Studies, ETH Zurich, 8092 Zurich, Switzerland}

\author{D.\ Malz}
\affiliation{Max Planck Institute of Quantum Optics, Hans-Kopfermann-Str.\ 1, D-85748 Garching, Germany}
\affiliation{Munich Quantum Center for Science and Technology, Schellingstr.\ 4, D-80799 München, Germany}
\affiliation{Department of Physics, Technische Universität München, James-Franck-Str.\ 1, 85748 Garching, Germany}

\author{J.\ I.\ Cirac}
\affiliation{Max Planck Institute of Quantum Optics, Hans-Kopfermann-Str.\ 1, D-85748 Garching, Germany}
\affiliation{Munich Quantum Center for Science and Technology, Schellingstr.\ 4, D-80799 München, Germany}

\begin{abstract}
Intermediate-scale quantum devices are becoming more reliable, and may soon be harnessed to solve useful computational tasks.
At the same time, common classical methods used to verify their computational output become intractable due to a prohibitive scaling of required resources with system size.
Inspired by recent experimental progress, here we describe and analyze efficient cross-platform verification protocols for quantum states and show how these can be used to verify computations.
We focus on the pair-wise comparison between distant nodes of a quantum network, identify the most promising protocols and then discuss how they can be implemented in laboratory settings.
As a proof of principle, we implement basic versions of these schemes on available quantum processors.
\end{abstract}

\maketitle

\section{Introduction}

As quantum computing platforms become more mature, it is essential to establish means to verify their computational output.
By construction, classical devices perform poorly at this task, especially as noisy intermediate-scale quantum (NISQ) devices \cite{Preskill2018} grow in size and may already be used to solve problems beyond the reach of classical machines \cite{Arute2019,Zhong2020}.
In recent years this has led to an increasing demand for efficient verification techniques beyond full state tomography \cite{Dariano2003,Liu2005,Cramer2010}.
By now there exists a range of different approaches to certify quantum states and computations, including techniques such as randomized benchmarking \cite{Knill2007}, fidelity estimation \cite{Flammia2011}, self-testing \cite{Reichardt2013} or interactive proofs \cite{Aharonov2017}.
While each approach has its context-specific advantages and disadvantages, a common goal is to find protocols with feasible resource requirements.
Several review articles have been put together in recent years on various aspects of quantum verification \cite{Gheorgiu2019,Eisert2020,Supic2020,Kliesch2021}.

One subset of approaches deals with quantum state verification, which can be used to test the reliability of quantum devices and their ability to generate specific states.
This is particularly suitable for NISQ devices, which cannot perform the deep circuits required for self-testing or interactive proofs.
The gold standard is to perform quantum state tomography (QST) to reconstruct the full quantum state, but this method requires a number of measurements that scales exponentially with the number of qubits.
Exponentially faster, but nevertheless still exponentially difficult, are fidelity estimation~\cite{Flammia2011,Silva2011,lanyon2017} and comparing randomized measurements~\cite{Elben2020,Elben2022}.
Forgoing theoretical guarantees, one could efficiently compare states via their classical shadow~\cite{Aaronson2018,Huang2020,Zhu2022}.

In these approaches, there is essentially no difference between comparing a quantum state to the predictions obtained from some classical description and comparing the measurement outcomes from two independent devices.
Inspired by recent experimental progress in realizing quantum local-area networks~\cite{Kurpiers2018,Magnard2020,Alshowkan2021}, we collect and assess quantum cross-platform verification schemes enabled by the presence of a quantum link between the devices.
In particular, a comparison of two unknown quantum states at distant locations may be performed using a combination of quantum state transfer \cite{Cirac1997} and the SWAP test \cite{Buhrman2001}, with or without additional ancilla qubits.
Unlike the aforementioned classical methods, the resource requirements of quantum cross-platform verification scale only linearly, which makes these schemes very attractive as devices become larger.
To identify the most useful scheme for near-term experiments, we explicitly account for the number of uses of the quantum channel, under the assumption that this is an expensive resource.

Building on these techniques for cross-platform verification of quantum states, we also present a set of efficient protocols for verifying quantum computations.
If two remote quantum machines are connected by a quantum channel, the cost of verifying computations inherits the linear scaling with system size from the state-verification schemes.
In particular, to find the process fidelity between two unitary operations up to an error $\epsilon$, the number of required samples scales as $O(1/\varepsilon^2)$, which follows from concentration bounds in probability theory.

\begin{figure*}
\centering
\includegraphics[width=\textwidth]{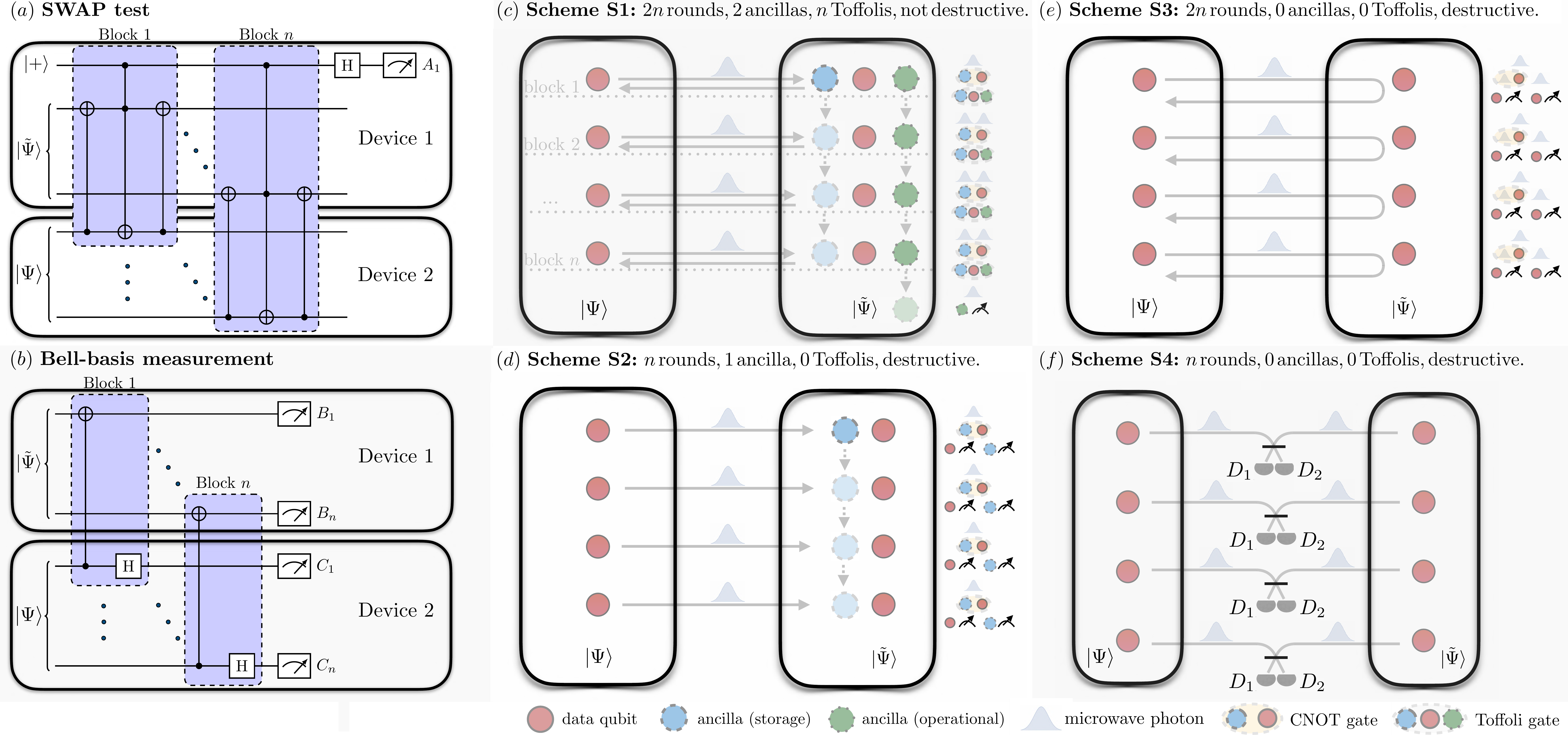}
    \caption{Protocols for state-overlap estimation in quantum networks.
    Quantum circuits for (\textit{a}) the SWAP test in terms of Toffoli gates using an ancilla qubit,
    (\textit{b}) the destructive overlap test with pairwise Bell-basis measurements.
    The latter requires additional classical post-processing.
    Repeated execution of both circuits yields estimate of $\mathcal{F} = |\langle\Psi|\tilde\Psi\rangle|^2$.
    (\textit{c}) \textbf{Scheme S1:} adaptation of standard SWAP test from (a) to sequential state comparison between two nodes of a network relying on auxiliary memory and operational qubits,
    (\textit{d}) \textbf{scheme S2:} adaptation of destructive overlap test from (b) to sequential state comparison between two nodes of a network relying on auxiliary memory qubits,
    (\textit{e}) \textbf{scheme S3:} adaptation of destructive test from (b) relying on a universal gate set for itinerant qubits,
    (\textit{f}) \textbf{scheme S4:} quantum optical realization of the destructive overlap test based on beam splitters and Hong-Ou-Mandel interference.
    }
    \label{fig:distributed_swap_schemes}
\end{figure*}

This article is structured as follows.
In Sec.~\ref{ssec:sequential-tests} we consider four distributed state-comparison schemes in quantum networks, discuss their gate count and necessary amount of communication and information transfer.
In Sec.~\ref{ssec:compare-computations} we
then discuss methods for comparing remote quantum computations by means of either state-comparison protocols, Hadamard tests or entanglement witnesses.
We complement this discussion by proof-of-principle demonstrations on single quantum devices, and discuss advantages and limitations of the distributed protocols.
In Sec.~\ref{sec:implementations} we highlight future prospects and possible implementations of cross-platform verification in quantum-network experiments, and provide realistic estimates for achievable fidelities.
Further promising directions for utilizing NISQ devices in verification experiments are mentioned in Sec.~\ref{sec:outlook}.

\section{Cross-platform verification \label{sec:cross-platform-verification}}

In this section we briefly review state-comparison protocols and adapt them to cross-platform verification in quantum networks, distinguishing between four distributed schemes in Sec.~\ref{ssec:sequential-tests}.
We conclude this subsection with a scaling analysis of the required number $m$ of experimental repetitions, \textit{i.e.}, sampling complexity, to obtain a $\epsilon$-close approximation to the fidelity between two states.
Using standard results from probability theory, $m = O(1/\epsilon^2)$ is controlled by the targeted precision of the estimate. 
Then we discuss three methods for comparing remote quantum computations in Sec.~\ref{ssec:compare-computations} and show that they inherit this beneficial scaling of resources from the state-comparison protocols of Sec.~\ref{ssec:sequential-tests}.
In Sec.~\ref{ssec:imperfections} we discuss imperfections and expected bottlenecks of the protocols.

\subsection{Comparing remote quantum states \label{ssec:sequential-tests}}

The fidelity $\mathcal{F}(\rho, \tilde \rho) \in [ 0, 1]$ is a measure of the similarity of two quantum states $\rho$ and $\tilde \rho$, which evaluates to one if and only if $\rho = \tilde \rho$ \cite{jozsa94}.
One common definition is given by
\begin{equation}\label{eq:fidelity}
    \mathcal{F}(\rho, \tilde \rho) = \frac{\mathrm{tr \left ( \rho \tilde \rho \right )}}{\max \left \{ \mathrm{tr}\left (\rho^2\right ), \mathrm{tr} \left ( {\tilde \rho}^2 \right ) \right \}}.
\end{equation}
Here we consider the scenario where $\rho$ and $\tilde \rho$ are generated at two distant locations, that are connected by a quantum channel over which quantum information can be transmitted with high fidelity.
As detailed in the following, the presence of such a quantum link enables the realization of efficient fidelity estimation procedures, based on distributed versions of the SWAP test.
The latter is a standard technique for estimating the trace overlap $\mathrm{tr}(\rho \tilde \rho)$ \cite{Buhrman2001}.
For simplicity, in the following we consider two pure $n$-qubit states $\ket{\Psi}$ and $\ket{\tilde \Psi}$ located at different nodes of a quantum network, in which case the trace overlap is equal to the fidelity $\mathcal{F}=|\langle\Psi|\tilde \Psi\rangle|^2$.
To apply the methods presented here to mixed states, one additionally needs to estimate the purities $\mathrm{tr}(\rho^2)$ and $\mathrm{tr}(\tilde \rho^2)$, \textit{e.g.}, with local copies of the states and a SWAP test, but this is beyond the scope of this manuscript.

The quantum circuit in Fig.~\ref{fig:distributed_swap_schemes}(a) describes the standard SWAP test to estimate $\hat{\mathcal{F}}$ using an ancillary qubit, upon repeated execution of the circuit.
The only measurement is performed on the ancilla, not on $\ket{\Psi}$ or $\ket{\tilde \Psi}$, which makes the test \textit{non-destructive}.
The Bell-basis measurement depicted in Fig.~\ref{fig:distributed_swap_schemes}(b) provides an ancilla-free alternative that requires fewer two-qubit gates, at the expense of destructive state measurements \cite{Escartin2013,Cincio_2018}.
In state-of-the-art implementations, it performs significantly better due to its reduced gate count and shorter circuit, as demonstrated on a superconducting (SC) quantum processor in App.~\ref{app:state-verification}.

Distributed versions of these two algorithms, between different nodes of a quantum network, may be performed by transmitting information over a quantum channel, and sequentially addressing pairs of qubits.
The following discussion is motivated by recent experimental achievements of deterministic state transfer of quantum information across a coherent quantum link \cite{Kurpiers2018,Magnard2020}.
Based on the well-established tests depicted in Fig.~\ref{fig:distributed_swap_schemes}(a) and Fig.~\ref{fig:distributed_swap_schemes}(b), here we propose and distinguish between four distributed protocols and compare their scaling of required resources with system size.
These four schemes are schematically depicted in Fig.~\ref{fig:distributed_swap_schemes}(c)-(f).
In all cases, their required channel usage scales linearly with system size, such that we assess their utility by inspecting the prefactors of this scaling and the number of required ancillas and gates.

\subsubsection{Scheme S1: Ancilla-based SWAP test}
The ancilla-based SWAP test \cite{Buhrman2001,Kang2019} may be executed sequentially by transmitting individual qubits, across a quantum channel, between two nodes of a network.
The circuit from Fig.~\ref{fig:distributed_swap_schemes}(a) can then be performed locally at one of the network nodes, which allows an efficient estimation of the overlap between two distant quantum states.
This does not require any additional classical communication.

\textit{Setup}.\textemdash
In Fig.~\ref{fig:distributed_swap_schemes}(c), we schematically depict two nodes of a quantum network with the $n$-qubit quantum states $\ket{\Psi}$ and $\ket{\tilde \Psi}$, respectively, given by
\begin{equation}\label{eq:states-1}
\ket{\overset{\smallbrl\smallsim\smallbrr}{\Psi}} = \sum_{i_1, ..., i_n} \overset{\smallbrl\smallsim\smallbrr}{c}_{i_1, ..., i_n} \ket{i_1, ..., i_n}.
\end{equation}

In this setup, $\hat{\mathcal{F}}$ can be estimated from a SWAP test as in Fig.~\ref{fig:distributed_swap_schemes}(a).
As shown in Fig.~\ref{fig:distributed_swap_schemes}(c), one of the local nodes holds two additional ancillary qubits.
One is a \textit{storage} qubit onto which the state of the first node is mapped sequentially.
The second ancilla is an \textit{operational} qubit that takes the role of the ancillary qubit from Fig.~\ref{fig:distributed_swap_schemes}(a).
It keeps track of the outcomes of the $n$ consecutive blocks of the SWAP test.
At the end of the protocol, it is measured in the computational basis, thus yielding an estimate for the fidelity $\hat{\mathcal F} = 2 m_0/m - 1$, with $m_0$ out of $m$ occurences of finding the operational ancilla in the $\ket{0}_\mathrm{op}$ state.

\renewcommand{\figurename}{PROT.}
\setcounter{figure}{0}

\begin{figure}[b]
\begin{protocol}{Distributed SWAP test}
\textit{Inputs.} $n$-qubit quantum states $\ket{\Psi}_\mathrm{A}$ and $\ket{\tilde \Psi}_\mathrm{B}$ in two different locations, $\mathrm{A}$ and $\mathrm{B}$.
Two ancillas at $\mathrm{B}$, one for $\mathrm{storage}$ and one $\mathrm{operational}$.\\
\sbline
\textit{Goal.} Estimate fidelity $\mathcal{F}(\ket{\Psi}_\mathrm{A}, \ket{\tilde \Psi}_\mathrm{B})$.
\sbline
\textit{The protocol.} Party $\mathrm{A}$ sends qubits to $\mathrm{B}$ and $\mathrm{B}$ sequentially addresses pairs of qubits in $n$ blocks.
Before and after the steps (1)-(3) are repeated $n$ times ($k = 1, ..., n$), $\mathrm{B}$ applies a $\mathrm{H}$ gate to $\mathrm{operational}$ qubit.
\begin{enumerate}
  \item \textbf{State transmission $\mathcal{T}_{A\rightarrow B}$.}
  \begin{enumerate}
    \item $\mathrm{A}$ transmits $k^\mathrm{th}$ qubit to $\mathrm{B}$.
    \item $\mathrm{B}$ stores received qubit in $\mathrm{storage}$ register.
  \end{enumerate}
  \item \textbf{Controlled-}$\mathrm{SWAP}$ operation.
  \begin{enumerate}
    \item $\mathrm{B}$ performs $\mathrm{cSWAP}$ on the $\mathrm{storage}$ qubit and their $k^\mathrm{th}$ qubit, conditioned on state of $\mathrm{operational}$ qubit.
  \end{enumerate}
  \item \textbf{State transmission $\mathcal{T}_{B\rightarrow A}$.}
  \begin{enumerate}
      \item $\mathrm{B}$ transmits $\mathrm{storage}$ qubit to $\mathrm{A}$.
      \item $\mathrm{A}$ stores received qubit and continues from step 1 until all qubits in the register have been sent and received.
  \end{enumerate}
    \item \textbf{Final measurement.}
        \begin{enumerate}
            \item $\mathrm{B}$ measures $\mathrm{operational}$ qubit in computational basis and counts occurences of $0$ and $1$ outcomes.
        The probability of finding the qubit in $0$ ($1$) is given by $(1+\mathcal{F})/2$ ($(1-\mathcal{F})/2$).
        \end{enumerate}
\end{enumerate}
\end{protocol}
\caption{Distributed SWAP test between two nodes of a network.}\label{fig:distributed-swap-procotol}
\end{figure}

\renewcommand{\figurename}{FIG.}
\setcounter{figure}{1}

\textit{State transfer}.\textemdash
Note that Eq.~\eqref{eq:states-1} can be rewritten as
\begin{equation}\label{eq:states-2}
\begin{aligned}
    \ket{\overset{\smallbrl\smallsim\smallbrr}{\Psi}} = & \overset{\smallbrl\smallsim\smallbrr}{\alpha}_1 \ket{0}_1 & \otimes & \sum_{i_2, ..., i_n} \overset{\smallbrl\smallsim\smallbrr}{a}_{i_2, ..., i_n} \ket{i_2, ..., i_n} \\
    + & 
    \overset{\smallbrl\smallsim\smallbrr}{\beta}_1 \ket{1}_1 & \otimes & \sum_{j_2, ..., j_n} \overset{\smallbrl\smallsim\smallbrr}{b}_{j_2, ..., j_n} \ket{j_2, ..., j_n}.
\end{aligned}
\end{equation}
As conceptually shown in Fig.~\ref{fig:distributed_swap_schemes}(c), the first step of the sequential SWAP test is a state transfer of the first qubit of the first node to the ancillary storage qubit, initially in $\ket{0}_\mathrm{st}$, of the second node,
\begin{equation}
\begin{aligned}
    \ket{\Psi} \otimes \ket{0}_\mathrm{st} \overset{\mathcal{T}}{\longrightarrow} &
    \sum_{i_2, ..., i_n} a_{i_2, ..., i_n}\ket{0,i_2,...,i_n} \otimes \alpha_1 \ket{0}_\mathrm{st}  \\
    + &  \sum_{i_2, ..., i_n} b_{i_2, ..., i_n}\ket{0,i_2,...,i_n} \otimes \beta_1 \ket{1}_\mathrm{st},
\end{aligned}
\end{equation}
Upon the state transfer denoted by $\mathcal{T}$, the first block of the circuit in Fig.~\ref{fig:distributed_swap_schemes}(a) may be performed locally at the receiving node.
The controlled-$\mathrm{SWAP}$ gate involves the two ancillary (\textit{storage} and \textit{operational}) qubits and the first qubit of the register $\ket{\tilde \Psi}$.
Hence, \textit{block 1} of the circuit in Fig.~\ref{fig:distributed_swap_schemes}(a) can be implemented as a controlled-$\mathrm{SWAP}$ ($\mathrm{cSWAP}$) gate acting on the first qubit of the second node and the \textit{storage} qubit, and conditioned on the \textit{operational} ancilla.
Subsequently the state transfer is reversed, implementing $\mathcal{T}^{^\dagger}$, and the \textit{storage} qubit is sent back to the first node.
This concludes $\textit{block 1}$ of the scheme S1 depicted in Fig.~\ref{fig:distributed_swap_schemes}(c).
Altogether this has achieved the operation
\begin{equation}\label{eq:operation-cswap}
\mathrm{cSWAP}_{\mathrm{op},1,\tilde 1} = \left ( \mathcal{T}_{\mathrm{st}\rightarrow 1} \right ) \circ \left ( \mathrm{cSWAP}_{\mathrm{op},\mathrm{st},\tilde 1} \right ) \circ \left ( \mathcal{T}_{1\rightarrow \mathrm{st}} \right ),
\end{equation}
where the $\mathrm{cSWAP}$ gate is exactly the operation performed in each block in Fig.~\ref{fig:distributed_swap_schemes}(a), conditioned on the first indexed qubit, and $``1"$ ($``\tilde 1"$) denotes the first qubit of the sending (receiving) register.
This gate-transfer block is repeated $n$ times, and thus the sequential SWAP test requires $2n$ rounds of quantum communication between the nodes.
In general, for two unequal states, this procedure leaves the states entangled.
For just estimating the state overlap, the reversed state transfer is not necessary, which reduces the rounds of communication by half.
A high-level summary of the full scheme is given in Prot.~\ref{fig:distributed-swap-procotol}, see also App.~\ref{app:sequential-tests} for further details.

\textit{Limitations}.\textemdash
The implementation of scheme S1 requires high-fidelity Toffoli gates.
Its minimum two-qubit gate count scales as $5n$ \cite{Smolin1996,Negkun2015}.
For arbitrary entangled states $\ket{\Psi}$ and $\ket{\tilde \Psi}$, it is not possible to interrupt (any of the discussed distributed schemes) after less than $O(n)$ rounds and get a good estimate $\hat{\mathcal{F}}(\ket{\Psi},\ket{\tilde \Psi})$.
The \textit{operational} ancilla qubit is only measured at the end of the last round.
The decoherence times of the operational ancilla and data qubits thus need to be at least of the order of the total runtime of the protocol.
The runtime of scheme S1 grows linearly in the number of qubits.

\subsubsection{Schemes S2 \& S3: Bell-basis measurement}
The Bell-basis measurement \cite{Escartin2013,Cincio_2018} in Fig.~\ref{fig:distributed_swap_schemes}(b) may be used to estimate the state overlap between two spatially separated quantum states in two distinct ways.
One scheme relies on the usage of an ancillary storage qubit, and the other works without ancillas.

\textit{With ancilla}.\textemdash
One instance of the destructive overlap test between two distant nodes connected by a quantum link is schematically depicted in Fig.~\ref{fig:distributed_swap_schemes}(d).
During the protocol, the quantum state $\ket{\Psi}$ at the sending node is sequentially transmitted to the second node.
Each qubit is first transmitted, stored in the ancillary \textit{storage} qubit, and finally measured together with the corresponding qubit from the receiving node.
In analogy to Eq.~\eqref{eq:operation-cswap} the first block of the circuit in Fig.~\ref{fig:distributed_swap_schemes}(b) can be decomposed as
\begin{equation}\label{eq:operation-block1}
\mathrm{Bell}_{1,\tilde 1} = \left ( \mathrm{H}_{\tilde 1} \circ \mathrm{CNOT}_{\tilde 1, \mathrm{st}} \right ) \circ \left ( \mathcal{T}_{1\rightarrow \mathrm{st}} \right ),
\end{equation}
with a controlled-NOT ($\mathrm{CNOT}$) gate between first qubit of the receiving register and the ancilla qubit, and a Hadamard gate $\mathrm{H}$ acting on the first qubit of the receiving register.
The two qubits are then measured in the computational basis, which altogether corresponds to a measurement in the Bell basis.
This block is repeated for all pairs of qubits.
The measurements can be performed after each block, cp.~Fig.~\ref{fig:distributed_swap_schemes}(b).
In contrast to scheme S1 introduced above, scheme S2 makes use of the quantum channel only $n$ times.
Accordingly, the scheme does not involve the reversed state transfer.
However it is destructive and the quantum states $\ket{\Psi}$ and $\ket{\tilde \Psi}$ cannot be recovered.

As shown in Fig.~\ref{fig:distributed_swap_schemes}(d), the Bell-basis measurement is performed by sequentially addressing pairs of qubits at the receiving node.
This requires one ancillary qubit, $n$ $\mathrm{CNOT}$ gates and $2n$ single-qubit measurements.
Scheme S2 thus requires much fewer resources than S1:
the five-fold overhead in two-qubit gate count and two-fold overhead in channel use makes it more suitable for near-term applications.
This is demonstrated in App.~\ref{app:state-verification} for a simple circuit executed on a (single) quantum processor, and underlined by the estimates provided in Sec.~\ref{sec:implementations}.
A step-by-step discussion of this scheme is included in App.~\ref{app:sequential-tests}.

\textit{Without ancilla}.\textemdash
Instead of storing the transmitted qubit at the receiving node using an ancillary storage qubit, the required two-qubit gates may also be implemented between flying and stationary qubits, \textit{e.g.}, based on photon-cavity scattering phases \cite{Duan2004}.
By reflecting a single-photon wave packet off of a cavity, whose frequency depends on a qubit that it is off-resonantly coupled with, a phase is imprinted on the photonic wavefunction.
Along these lines, recent work has demonstrated the implementation of universal quantum gates for traveling microwave photons \cite{Reuer2022}, where gates are enabled by controllable couplings between SC qubits and itinerant microwave fields.

A deterministic and non-postselected controlled-PHASE ($\mathrm{CPHASE}$) gate between two distant qubits may be realized by mapping one qubit onto a flying qubit, sending it over to the second node, imprinting a state-dependent scattering phase on the photon and transmitting it back to the first node.
This can be extended to a $\mathrm{CNOT}$ gate between distant qubits, as the latter can be implemented using a $\mathrm{CPHASE}$ and few single-qubit gates at both nodes.
A faithful implementation of scheme S3 requires high-fidelity two-qubit gates between a traveling qubit and a stationary qubit, and while current fidelities are rather low, they can be expected to increase significantly in the future.
The resource requirements of scheme S3 are similar to scheme S2, but it needs $2n$ rounds of communication.
However, it does not require any ancillary qubits.

\textit{Limitations}.\textemdash
The decision rule of the ancilla-free test is based on the parity of a classical bitstring [see below and App.~\ref{app:state-verification}], which is very sensitive to bit-flip errors.
This may lead to misclassification of measurement results, which we discuss further in Sec.~\ref{ssec:imperfections}.
This equally affects schemes S2, S3 and S4.

\subsubsection{Scheme S4: Hong-Ou-Mandel interference \label{ssec:s4-hom}}
The Hong-Ou-Mandel (HOM) effect occurs when two indistinguishable particles enter and interfere at a symmetric beamsplitter, and the output state depends on the symmetry of the two-particle wavefunction \cite{Hong1987}.
As shown in Refs.~\cite{Horn2005,Escartin2013} the Bell-basis measurement from Fig.~\ref{fig:distributed_swap_schemes}(b) is formally equivalent to the HOM effect from quantum optics.

Previous experimental progress has enabled deterministic shaping and controlling of photonic wavefunctions both in the optical and microwave-frequency domains \cite{Pechal2014,Morin2019}, and stationary qubits have recently been mapped onto traveling single-photon states with high fidelity using tools from cavity quantum electrodynamics \cite{Kurpiers2018}.
This motivates the search for an all-optical protocol for estimating the overlap between quantum states at distant locations.

An optical discrimination of two photonic states may be achieved by directing two photons into separate input ports of a symmetric beam splitter and recording counts from two photodetectors located at the two output ports.
The shape of a single-photon state
\begin{equation}
    \ket{1} = \int \mathrm{d} t \ \psi(t) \hat a_t^\dagger \ket{\mathrm{vac}}
\end{equation}
is described by its temporal mode function $\psi(t)$, with the photon mode creation operator $a^\dagger_t$ at time $t$.
When being sent into the input ports of a symmetric beamsplitter, the probability of finding two incoming photon states $\psi_1(t)$ and $\psi_2(t)$ at the same output port can be calculated from their pair correlation function and is related to their temporal overlap \cite{Loudon,Woolley2013}.
It can be obtained as a function of experimental parameters, such as the relative frequency difference and the photons' relative delay in arrival time $\delta t$, see also App.~\ref{app:sequential-tests}.

Time-bin qubits are described by wavefunctions with non-overlapping supports \cite{Brendel1999,Brecht2015}.
In recent experiments, time-bin superpositions of two propagating temporal microwave modes have been used to transfer the state of a stationary SC qubit over a macroscopic distance \cite{Kurpiers2018}.
This makes these states ideal candidates for an experimental realization of a distributed overlap test using microwave photons.
The quantum states $\ket{\Psi}$ and $\ket{\tilde \Psi}$ from Eq.~\eqref{eq:states-1} and the schematic in Fig.~\ref{fig:distributed_swap_schemes} may be sequentially sent to a beamsplitter by mapping the stationary qubits onto traveling time-bin encoded qubits,
\begin{equation}
\begin{aligned}
    \ket{\overset{\smallbrl\smallsim\smallbrr}{\Psi}} \otimes \ket{\mathrm{vac}} \longrightarrow &
    \sum_{i_2, ..., i_n} \overset{\smallbrl\smallsim\smallbrr}{a}_{i_2, ..., i_n}\ket{0,i_2,...,i_n} \otimes \overset{\smallbrl\smallsim\smallbrr}{\alpha}_1 \ket{0}_\mathrm{\overset{\smallbrl\smallsim\smallbrr}{bin}}  \\
    + &  \sum_{i_2, ..., i_n} \overset{\smallbrl\smallsim\smallbrr}{b}_{i_2, ..., i_n}\ket{0,i_2,...,i_n} \otimes \overset{\smallbrl\smallsim\smallbrr}{\beta}_1 \ket{1}_\mathrm{\overset{\smallbrl\smallsim\smallbrr}{bin}},
\end{aligned}
\end{equation}
where $\overset{\smallbrl\smallsim\smallbrr}{\mathrm{bin}}$ denotes a distinct photonic mode, and $\ket{0}$ ($\ket{1}$) denotes the absence (presence) of a photon in each time bin.
A HOM interference experiment on the two time-bin qubits from both nodes realizes an optical Bell-basis measurement. 
The pair-wise comparison between modes $\mathrm{bin}$ and $\overset{\sim}{\mathrm{bin}}$ passes the test if both photons are detected at the same output port, and the test fails if there is a coincidence count \cite{Escartin2013}.
In this fashion, the two registers may be mapped onto temporal photonic modes and sequentially compared in $n$ runs.

\subsubsection{Fidelity estimation}

The previously introduced schemes are distributed versions of established methods for estimating the closeness of quantum states, as quantified by $\mathrm{tr}(\rho \tilde \rho)$ for two arbitrary mixed states $\rho$ and $\tilde \rho$, which reduces to $|\langle \Psi | \tilde \Psi \rangle |^2$ for two pure states $\ket{\Psi}$ and $\ket{\tilde \Psi}$.
From both the standard SWAP test [Fig.~\ref{fig:distributed_swap_schemes}(a)] and the destructive overlap test [Fig.~\ref{fig:distributed_swap_schemes}(b)], the fidelity $\mathcal{F}$ can be estimated and confidence intervals derived using statistical methods and the framework of hypothesis testing \cite{Yu2022}.

In each experimental realization, a scheme-dependent criterion decides whether two states \textit{pass} or \textit{fail} the test:
(\textit{i}) the criterion for the standard SWAP test simply assigns these labels to outcomes of a computational basis measurement, \textit{i.e.}, they pass (fail) if the outcome is $A_1 = 0$ ($A_1 = 1$) in Fig.~\ref{fig:distributed_swap_schemes}(a);
(\textit{ii}) the criterion for the destructive test is slightly more involved and assigns the labels to the parity of the bitwise $\mathrm{AND}$ of the bitstrings $B_1 ... B_n$ and $C_1 ... C_n$ in Fig.~\ref{fig:distributed_swap_schemes}(b), \textit{i.e.}, they pass (fail) the test if the parity is even (odd) \cite{Cincio_2018}, cf.~also App.~\ref{app:state-verification}.

In both cases, the fidelity can be estimated upon $m$ repetitions from the number of passed tests $m_\mathrm{p}$ as $\hat{\mathcal{F}} = 2 m_\mathrm{p}/m - 1$.
Using Hoeffding's inequality \cite{hoeffding63}, the probability that $\hat{\mathcal{F}}$ deviates from the true fidelity $\mathcal{F}$ by more than $\epsilon$ is bounded by
\begin{equation}\label{eq:hoeffding}
\mathrm{Pr}\left ( |\hat{\mathcal{F}} - \mathcal{F}| \geq \epsilon \right ) \leq 2 e^{-m\epsilon^2/2}.
\end{equation}
From this we obtain a $(1-\alpha)$ confidence interval $[\hat{\mathcal{F}}-\epsilon, \hat{\mathcal{F}}+\epsilon]$ by repeating one of the above protocols $m = \Theta \left ( \ln(1/\alpha)/\epsilon^2 \right )$ times.
The designated confidence level $(1-\alpha)$ refers to the probability that the confidence interval contains the true fidelity.
Typical choices are $\alpha = 0.01$ or $\alpha=0.05$.
Hence, the cost of gaining confidence is sublinear in terms of confidence level and quadratic in terms of precision and does not depend on system size $n$.
While this result yields a simple scaling, tighter bounds can be derived for binomial proportions, cf.~App.~\ref{app:confidence} for more details.

\subsection{Comparing remote quantum computations \label{ssec:compare-computations}}

The previous schemes are targeted at comparing distant quantum states, but they may also be utilized for discriminating quantum computations efficiently.
We present three ways of comparing quantum operations, given that a method for comparing quantum states is available.
We are interested in the case where two devices each perform a unitary quantum operation, $U_\mathrm{L}$ and $U_\mathrm{R}$, respectively, and the goal is to compare those operations with as few measurements as possible.

\begin{figure}
\centering
\includegraphics[width=0.482\textwidth]{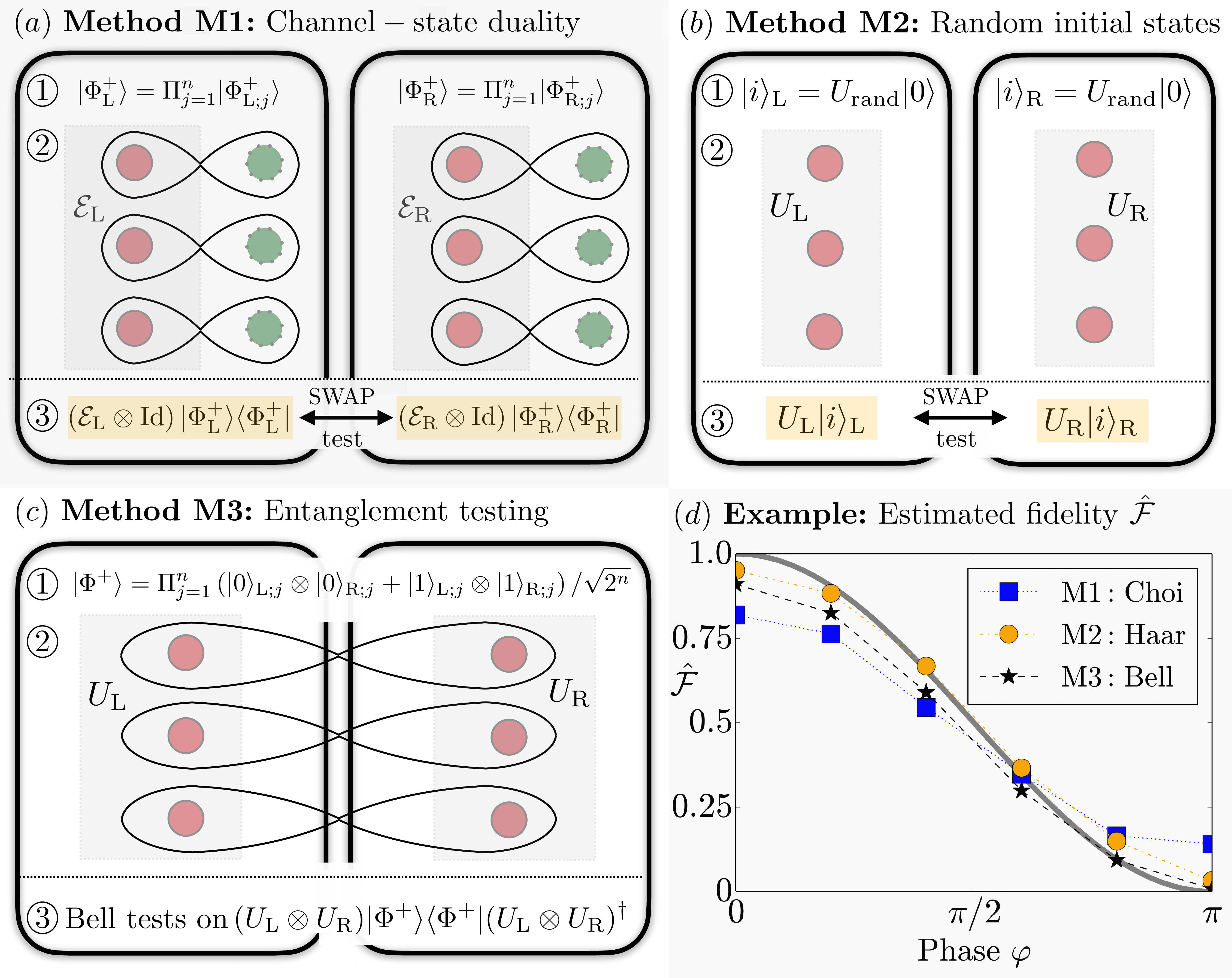}
    \caption{Protocols for comparing quantum computations performed at different nodes of a quantum network.
    (a) \textbf{Method M1:} Distributed overlap estimation of Choi states of two quantum channels,
    (b) \textbf{method M2:} fidelity estimation for a set of random initial states,
    (c) \textbf{method M3:} Bell tests on initially maximally entangled state onto which unitaries have been appled on each device separately.
    (d) Example: comparison of two single-qubit unitaries $U_\mathrm{L} = \mathrm{H}$ and $U_\mathrm{R} = \mathrm{P}(\varphi) \mathrm{H}$, run on IBM superconducting quantum processor using $m = 16384$ shots for each circuit.
    Bell-basis measurements and classical post-processing are used for fidelity estimation.
    $\mathrm{H}$ and $\mathrm{P}$ denote Hadamard and phase gate, respectively.
    The gray curve corresponds to the analytical result $\mathcal{F} = (1+\cos\varphi)/2$.
    Statistical error bars are smaller than labels in this plot.
    }
    \label{fig:compare-comps}
\end{figure}

In the following discussion we assign the two devices the labels $\sigma = \mathrm{L}$ (\textit{left} device) and $\sigma = \mathrm{R}$ (\textit{right} device), cf.~Fig.~\ref{fig:compare-comps}.
As an illustrative example, we put the three methods into practice by running them on a single SC quantum processor and comparing their performance.
For the methods that rely on state comparison, we utilize the Bell-basis measurements from Fig.~\ref{fig:distributed_swap_schemes}(b).
While this simple demonstration does not capture the distributed or sequential nature of the protocols, it serves as an illustration of the differences in performance on a single state-of-the-art device.
In the following, $n$ denotes the number of qubits per device and $d = 2^n$ is the local Hilbert space dimension.

\subsubsection{Method M1: Channel-state duality}
The Choi-Jamio{\l}kowski isomorphism establishes a correspondence between quantum channels, as described by completely positive maps, and quantum states, as described by density matrices \cite{Jami1972,Choi1975}.
It thus enables to map statements about channels to statements about states.
To compare two quantum operations each described by a quantum channel, one may compare their respective Choi states \cite{xie2022}.
This requires the preparation of a maximally entangled state, \textit{e.g.}, $\ket{\Phi_\sigma^+} = (1/\sqrt{d}) \sum_{i=0}^{d-1} \ket{i^\sigma_{\mathrm{c}}} \otimes \ket{i^\sigma_\mathrm{a}}$, between the computational (subscript ``$\mathrm{c}$") and an ancillary (``$\mathrm{a}$") register on each of the two devices ($\sigma = \mathrm{L}, \mathrm{R}$).
The quantum channel $\mathcal{E}_\sigma$ is applied to the computational register of $\ket{\Phi^+_\sigma}$, which yields the two states
$\rho_\sigma = (1/d) \sum_{i,j} \mathcal{E}_\sigma (\ket{i^\sigma_\mathrm{c}}\bra{j^\sigma_\mathrm{c}}) \otimes \ket{i^\sigma_\mathrm{a}}\bra{j^\sigma_\mathrm{a}}$, for $\sigma = \mathrm{L}, \mathrm{R}$.
Here we consider unitary quantum channels such that $\mathcal{E}_\sigma(\rho) = U_\sigma^\dagger \rho U_\sigma$.

Upon preparation of the two Choi states $\rho_\mathrm{L}$ and $\rho_\mathrm{R}$, one of the protocols from Sec.~\ref{ssec:sequential-tests} can be used to estimate their overlap $\mathrm{tr} \left(\rho_\mathrm{L} \rho_\mathrm{R}\right)$, which quantifies the process fidelity between $\ul$ and $\ur$.
In analogy to the fidelity between two quantum states, the latter compares the quantum operations performed on both devices \cite{Ragkinsky01}.
For any unitaries $U_\mathrm{L}$ and $U_\mathrm{R}$, it follows that the process fidelity
\begin{equation}\label{eq:process-fidelity}
    \mathcal{F}_\mathrm{p}(U_\mathrm{L},U_\mathrm{R}) := \frac{|\mathrm{tr}(U_\mathrm{L}^\dagger U_\mathrm{R})|^2}{d^2}
\end{equation}
equals the Choi-state overlap, \textit{i.e.}, $\mathcal{F}_\mathrm{p} = \mathrm{tr}\left( \rho_\mathrm{L} \rho_\mathrm{R} \right)$.
Since $0 \leq \mathcal{F}_\mathrm{p} \leq 1$ and $\mathcal{F}_\mathrm{p} = 1$ if and only if $U_\mathrm{L} = U_\mathrm{R}$, any of the previously discussed overlap tests, performed on the Choi states, checks whether both devices have reliably run the same computation.
The fidelity can then be estimated as before, \textit{i.e.}, $\hat{\mathcal{F}}_\mathrm{p} = 2m_\mathrm{p}/m - 1$, up to an error $\epsilon$ and confidence ($1-\alpha)$ can be gained efficiently by repeating the test $m = O(\ln(1/\alpha)/\epsilon^2)$ times.

This method is conceptually simple, but requires the preparation of maximally entangled states, and a two-fold overhead in uses of the quantum channel for comparing $2n$-qubit states instead of $n$-qubit states.
When applied to a simple example, on four qubits of a single quantum processor in which two qubits each mimic one of the two devices, the performance is the worst of all three methods M1-M3, cf.~Fig.~\ref{fig:compare-comps}(d).
More details are provided in App.~\ref{app:comp-state-three-methods}.

\subsubsection{Method M2: Sampling initial states}

Another set of strategies for comparing unitary operations between devices relies on random sampling of initial states, applying the unitaries on each device and comparing the resulting states using one of the above schemes.
Averaging over the outcome for a few initial states yields an estimate for the similarity of the two unitaries $\ul$ and $\ur$.
Here we distinguish between three different instances of this approach.
In contrast to methods M1 and M3, these approaches (except for one) do not require the preparation of entangled states as an additional resource.

We first consider a method to estimate Eq.~\eqref{eq:process-fidelity} using one of the above state-comparison protocols, which is based on sampling Haar-random initial states.
Then we utilize the so-called Hadamard test to obtain an efficient estimate of $\mathrm{tr}(\uld \ur)/d = \sum_p \langle p | \uld \ur | p\rangle/d$ by sampling a few states $\ket{p}$ from a complete basis.
The Hadamard test is different from the SWAP test and Bell-basis measurements shown in Fig.~\ref{fig:distributed_swap_schemes} and is briefly reviewed in App.~\ref{app:comp-state-three-methods}.
Finally, we discuss a method that is, again, based on overlap estimation via SWAP tests or Bell-basis measurements [Sec.~\ref{ssec:sequential-tests}] and provides access to an overlap measure different from the process fidelity in Eq.~\eqref{eq:process-fidelity}.
For current experimental implementations, as we will see, the latter is the most straightforward one to realize.

\textit{(i) Sampling from unitary 2-designs}.\textemdash
The process fidelity defined in Eq.~\eqref{eq:process-fidelity} can also be expressed as \cite{horodecki99,nielsen01}
\begin{equation}
\mathcal{F}_\mathrm{p}(\ul, \ur) = \frac{(d+1) \mathcal{F}_\mathrm{av}(\ul, \ur) - 1}{d},
\end{equation}
in terms of an \textit{average} fidelity, which can be obtained as an average over a finite set of initial states,
\begin{equation}\label{eq:average-fidelity}
\mathcal{F}_\mathrm{av}(\ul,\ur) = \frac{1}{K} \sum_{k=1}^K |\langle k| \uld \ur | k \rangle|^2,
\end{equation}
with $|k\rangle=U_k|0\rangle$.
The unitaries $U_1, ..., U_K$ must form a 2-design, which can be approximated by random quantum circuits \cite{brandao16}.

The sum in Eq.~\eqref{eq:average-fidelity} contains the state fidelities $\mathcal{F}(\ul |k\rangle, \ur |k\rangle)$ which are estimated using one of the schemes from above, separately for each initial state $|k\rangle$, cf.~Fig.~\ref{fig:compare-comps}(b).
Denoting the success probability of passing the $k$th test by $p_k = m_\mathrm{p}^{(k)}/m^{(k)}$, the process fidelity can be estimated as $\hat{\mathcal{F}}_\mathrm{p} = 2(d+1)/(Ld) \sum_{k=1}^L p_k - (d+2)/d$ using $L$ initial states $U_k \ket{0} \ (k = 1, ..., L)$ sampled from $\{ U_k \}_{k \in \{1, ..., K\}}$ with replacement.

In Fig.~\ref{fig:compare-comps}(d) it is exemplarily shown that this approach performs significantly better than method M1, when comparing two single-qubit gates on a single device.
In this simple example, we estimate all terms in Eq.~\eqref{eq:average-fidelity} by repeating the experiment for all $K = 24$ single-qubit Clifford gates.
For larger systems, $\mathcal{F}_\mathrm{av}$ can be estimated from a few randomly selected initial states, and $O(1/\epsilon^2)$ repetitions per state to achieve a $\epsilon$-close approximation of each term in the sum of Eq.~\eqref{eq:average-fidelity}.

\begin{figure}
\centering
\includegraphics[width=0.35\textwidth]{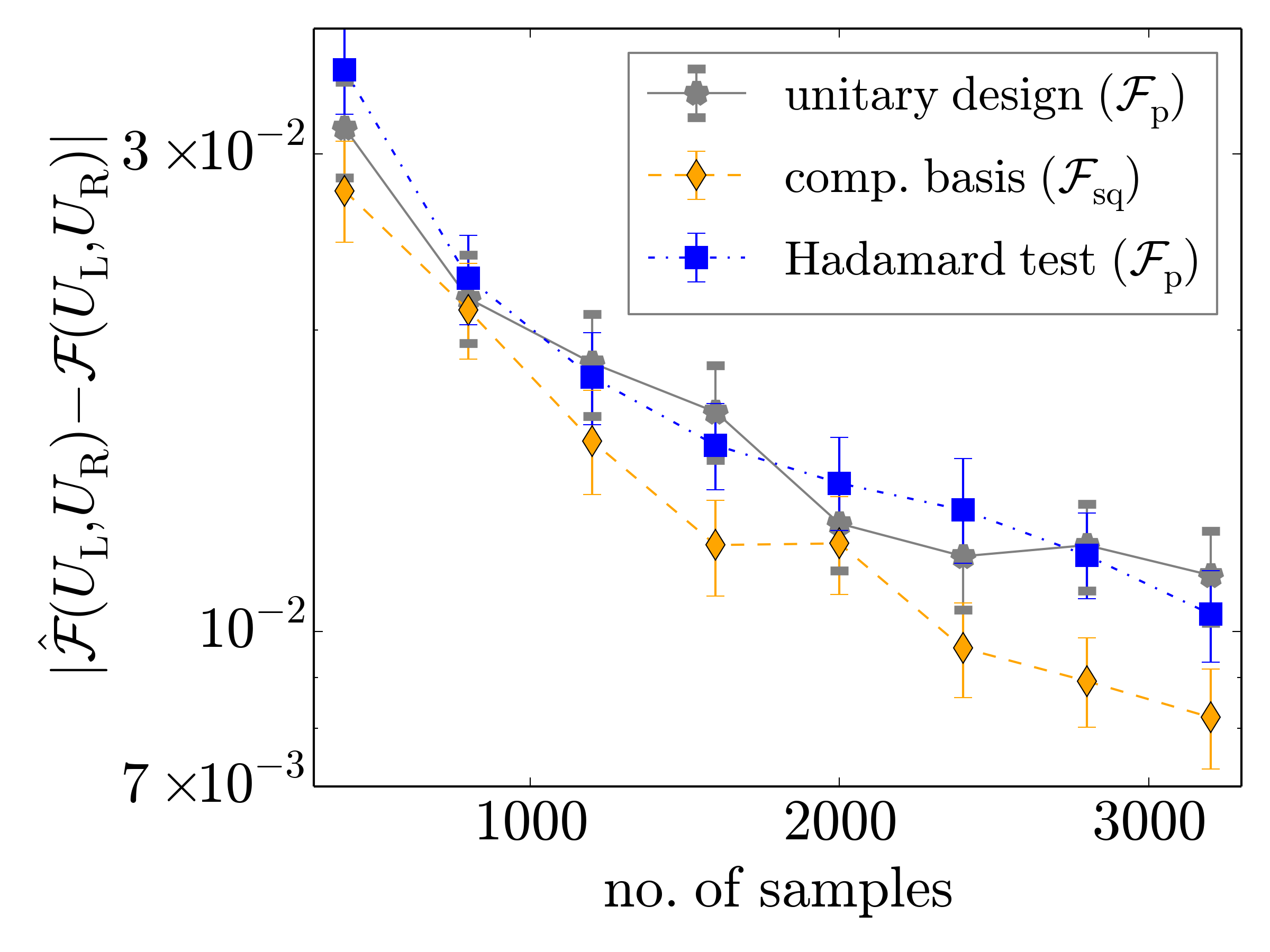}
    \caption{Numerical results from fidelity estimation using the different strategies outlined under the section Sec.~\ref{ssec:compare-computations} (\textit{Method M2}).
    The estimates are obtained from simulating a SWAP test for a random $5$-qubit unitary $\ul$ and $\ur = M \ul$, where $M$ is a product of random single-qubit rotations.
    The total number of samples $m = m_s m_b$ is plotted against the mean value of the absolute difference between the estimated and actual fidelities.
    For each initial state we execute the circuit $m_s = 100$ times and collect statistics by sampling $m_b \in \{ 4, 8, ..., 32 \}$ initial states.
    Error bars are obtained from bootstrap sampling.
    }
    \label{fig:compare-sample-complexity}
\end{figure}

In Fig.~\ref{fig:compare-sample-complexity} we show a numerical example, where we estimate $\hat{\mathcal{F}}_\mathrm{p}$ and show the deviation from the true value as a function of total number of circuit executions.
For this we numerically simulate a SWAP test to compare a random unitary $U$ with itself, but the second unitary includes additional random single-qubit rotations of all $n = 5$ qubits, cf.~App.~\ref{app:comp-state-three-methods} for further details.

This method requires sampling $n$-qubit Clifford operators uniformly at random, and running one of the state-comparison protocols for each of the randomly drawn initial states.
In an experiment with distant machines, using one of the schemes S2 or S4, the required number of qubit transmissions per experiment is $n$.

\textit{(ii) Modified Hadamard tests}.\textemdash
The Hadamard test is a standard method in quantum computation to create random variables whose expectation values are $\mathrm{Re}~\langle \psi | U | \psi \rangle$ and $\mathrm{Im}~\langle \psi | U | \psi \rangle$, respectively \cite{Aharonov2009}.
It can be adapted to estimate $\langle \psi | \uld \ur | \psi \rangle$, cf.~App.~\ref{app:comp-state-three-methods} for details.
We consider two ways of utilizing this test for estimating $\mathrm{tr} (\uld \ur)/d$ and its absolute square, $\mathcal{F}_p(\ul, \ur)$.

In the first one we measure $\langle p|\uld \ur|p\rangle$ for a random vector $p$ randomly drawn from a probability distribution $\mathcal{D}$ such that $\mathbb{E}_{p \sim \mathcal{D}} [\langle p | \uld \ur | p \rangle ] = \mathrm{tr} ( \uld \ur )/d$.
This is equivalent to sampling $\ket{p}$ from the computational basis, \textit{i.e.}, $p \in \{ 0, 1 \}^n$ \cite{Shor2007,Chen2022}.
Up to an additive error $\epsilon$,  the normalized trace $\mathrm{tr}(\uld \ur)/d$ can be efficiently estimated independent of system size, by sampling $m_\mathrm{b} = O(\ln(1/\alpha)/\epsilon^2)$ basis states.
A numerical example for the required number of samples is shown in Fig.~\ref{fig:compare-sample-complexity}, where the two compared unitaries differ only by random single-qubit rotations.

Another way of using the Hadamard test to compare $\ul$ and $\ur$ is by preparing a maximally entangled state between both registers, and applying controlled unitaries $\mathrm{C}\ul$ and $\mathrm{C}\ur$ with an ancilla acting as a control.
This way, we can estimate $\mathrm{tr}(\ul^* \ur)/d$ using $O(\ln(1/\alpha)/\epsilon^2)$ repetitions of the experiment.
If the devices were to implement $\ul = U$ and $\ur = U^T$, for a given unitary $U$, this allows to measure $ |\mathrm{tr}(U^\dagger U)|^2/d^2 = \mathcal{F}_\mathrm{p}(U,U) = 1$ and deviations from a reliable implementation of $U$ and $U^T$ may be detected by finding a process fidelity smaller than one.

These methods both require the implementation of controlled unitaries at both devices.
In the first one we need to transmit $(n+1)$ qubits between the devices, and in the second one we need to prepare a maximally entangled state at the beginning, and transmit only one qubit later.

\textit{(iii) Sampling computational basis states}.\textemdash
While the previous sampling algorithms enable efficient comparison of two unitaries based on the metric $\mathcal{F}_\mathrm{p}$ defined in Eq.~\eqref{eq:process-fidelity}, here we mention a conceptually simpler method that allows to efficiently obtain an estimate of
\begin{equation}\label{eq:fid-sq-maintext}
    \mathcal{F}_\mathrm{sq} := \frac{1}{d} \sum_{p \in \{0,1\}^n} |\langle p | U_L^\dagger U_R | p \rangle |^2,
\end{equation}
which is one if and only if $\ul = \ur$.
To measure each term in this sum, it is sufficient to prepare the two registers in a product state and perform one of the state-comparison protocols.
We estimate $\hat{\mathcal{F}}_\mathrm{sq} = \sum_{k=1}^{m_\mathrm{b}} (1/m_\mathrm{b}) |\langle p_k | \uld \ur | p_k \rangle |^2$, drawing $m_\mathrm{b}$ basis states uniformly at random from the computational basis.
We obtain an estimate of $\mathcal{F}_\mathrm{sq}$, up to an additive error $\epsilon$, by sampling $m_\mathrm{b} = O(1/\varepsilon^2)$ states.
An example of the performance of this method, for the case of random single-qubit rotations considered before, is shown in Fig.~\ref{fig:compare-sample-complexity}.
The number of necessary single-qubit transmissions is $n$, using either of the schemes S2 or S4.
This approach is the most suitable for current hardware, as the comparison of two unitaries reduces to the comparison of states [Sec.~\ref{ssec:sequential-tests}], for a few randomly selected states that are easy to prepare.

\subsubsection{Method M3: Entanglement testing}
The comparison of two remote computations may also be performed by running a series of entanglement tests.
To this end, the devices may be prepared in an entangled state, such that the entanglement is distributed across the devices, cf.~Fig.~\ref{fig:compare-comps}(c).
Suppose they are prepared in a maximally entangled state, \textit{e.g.}, $\ket{\Phi^+} = (1/\sqrt{d}) \sum_{i=0}^{d-1} \ket{i^\mathrm{L}} \otimes \ket{i^\mathrm{R}}$.
Since $(U \otimes \mathbb{1}) |\Phi^+\rangle = (\mathbb{1} \otimes U^T) |\Phi^+\rangle$, the state is invariant under transformations of the form $U\otimes U^*$.

If $U_\mathrm{L}$ is applied on the \textit{left} device, the full state will remain maximally entangled if and only if $U_\mathrm{R} = U_\mathrm{L}^*$ is applied on the \textit{right} device.
Hence if $U_\mathrm{L} = U$ and $U_\mathrm{R} = U^*$ are perfectly implemented, the CHSH test statistic $\langle S \rangle = 2 \sqrt{2}$ saturates Tsirelson's bound, and imperfections may be detected by deviations from this value.
The implemented quantum operations on both devices may thus be compared by testing the amount of entanglement that remains in the final state.

In Fig.~\ref{fig:compare-comps}(d) we demonstrate this for the same single-qubit example that we used already for the previous methods.
We perform a CHSH experiment on a SC quantum processor and compare the outcomes with the analytical result $\langle S \rangle = 2\sqrt{2} \mathcal{F}$, cf.~App.~\ref{app:comp-state-three-methods}, finding reasonably good agreement.
In practice, the main bottleneck of this approach is the preparation of a maximally entangled state between the two devices.
However, it does not require any additional quantum state transfer.

\subsection{Imperfections \label{ssec:imperfections}}
In the protocols above, we may distinguish between two kinds of imperfections, \textit{i.e.}, (\textit{i}) local errors on individual nodes, which affect schemes S1 and S2 most severely, and (\textit{ii}) errors that occur due to imperfect state transfer.
Here we provide simple examples and map out typical errors of the first kind on superconducting and trapped-ion quantum processors.
They lead to false positives and negatives in the classification of \textit{pass} and \textit{fail} outcomes in the fidelity-estimation procedure.
Then we briefly give a perspective on imperfect state transfer and how it limits the implementation of the protocols.

\textit{Local errors}.\textemdash
Imperfect gates and finite relaxation times may be characterized for each node separately.
Randomized benchmarking can be used to estimate average error rates under the implementation of random gate sequences \cite{Helsen2022}.
Sources of error include measurement and qubit errors (relaxation and dephasing), finite gate durations and gate errors, but also crosstalk between gates.
The latter is often dominated by non-local and parallelized gates.
The ancilla-free test from Fig.~\ref{fig:distributed_swap_schemes}(b) may involve the parallel execution of $\mathrm{CNOT}$ gates and provide a blueprint of such errors.

\begin{figure}
\centering
\includegraphics[width=0.46\textwidth]{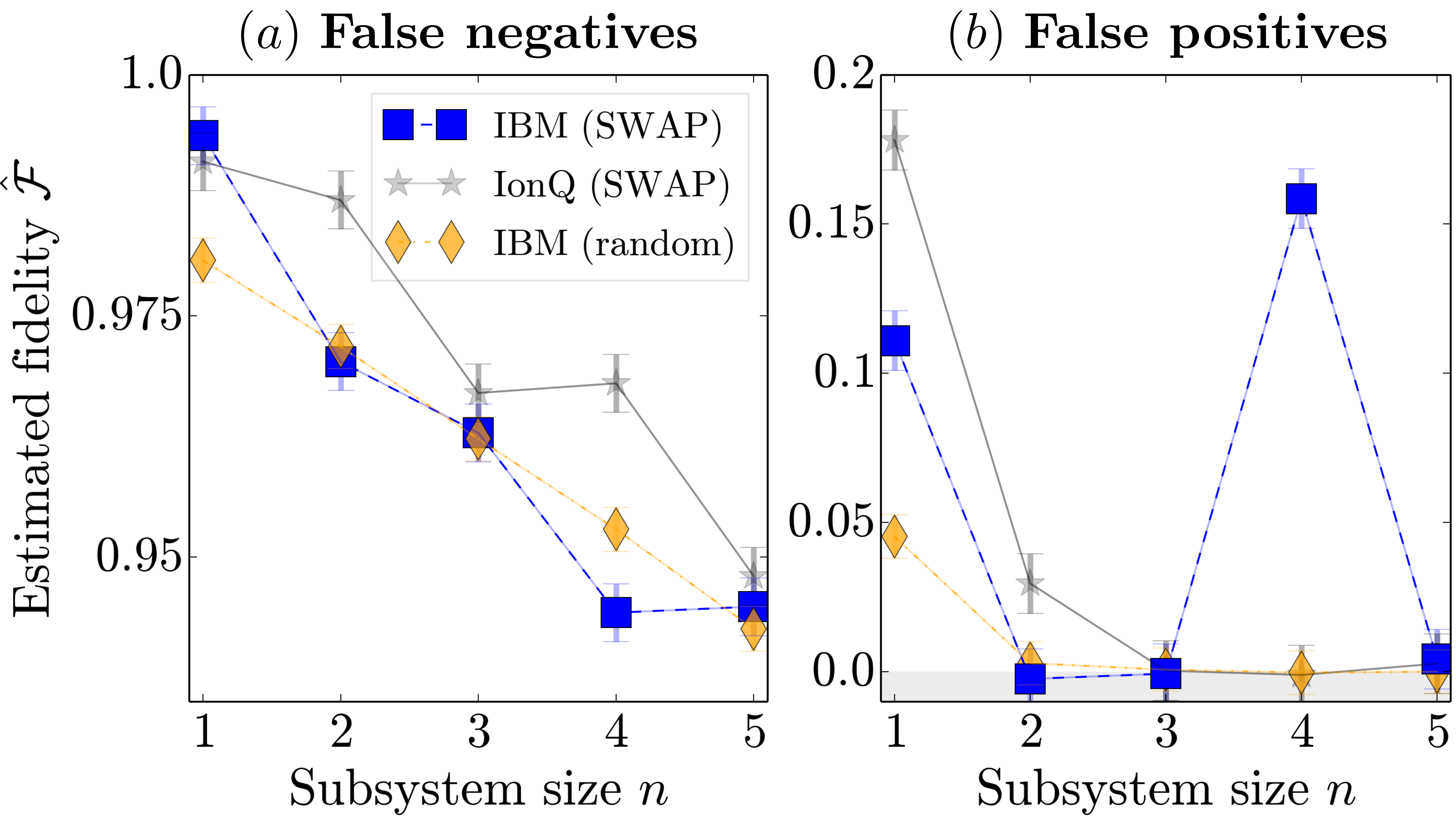}
    \caption{Estimated fidelity $\hat{\mathcal{F}}(\ket{\Psi}, \ket{\tilde \Psi})$ for initially prepared states (a) $\ket{\Psi} = \ket{\tilde \Psi}  = |+\rangle^{\otimes n}$ and (b) $|\Psi\rangle = |+\rangle^{\otimes n}$, $|\tilde \Psi\rangle = |-\rangle^{\otimes n}$, respectively, as a function of number of qubits $n$.
    Circuit shown in Fig.~\ref{fig:distributed_swap_schemes}(b) performed on superconducting (IBM, denoted by blue squares) and trapped-ion (IonQ, denoted by gray stars) quantum processors.
    In the legend it is denoted by ``$\mathrm{SWAP}$" for brevity.
    Another estimate is obtained from randomized measurements (orange diamonds) on the same SC device.
    Statistical error bars are given for each data point.
    Each circuit is executed in $65536$ times.
    False-negative and false-positive rates are given by $1-\hat{\mathcal{F}}$ in (a) and $\hat{\mathcal{F}}$ in (b), respectively.
    }
    \label{fig:benchmarking}
\end{figure}

In Fig.~\ref{fig:benchmarking} we show the estimated fidelity for target states that are (a) identical ($\mathcal{F} = 1$) and (b) orthogonal ($\mathcal{F} = 0$), respectively.
We employ two methods to estimate the fidelity between initially prepared states, \textit{i.e.}, Bell-basis measurements and randomized measurements (see Ref.~\cite{Elben2020} for further details).
For both methods, we compare two subsystems with $n = (1-5)$ qubits, and naturally find that the fidelity tends to decrease as a function of system size due to qubit, gate and measurement errors.

False negatives, \textit{i.e.}, the erroneous declaration of \textit{failed} test runs when they should ideally yield only \textit{pass} instances for identical target states, are depicted in Fig.~\ref{fig:benchmarking}(a).
For both architectures the estimated fidelity is consistent with an exponential decay.
However the exact dependence on system size is slightly more involved and depends on the particular device.
The non-monotonous behaviour may be due to crosstalk errors.
Using randomized measurements, the estimated fidelity is closer to a simple exponential decay.

False positives, \textit{i.e.}, the erroneous declaration of \textit{passed} test runs when they should ideally yield only \textit{fail} instances for orthogonal target states, are depicted in Fig.~\ref{fig:benchmarking}(b).
The estimated fidelities are close to zero, as expected, except for some outliers at particular system sizes.
This may again be attributed to non-local errors.
If it were caused mainly by decoherence and other sources of gate infidelity, it should equally affect the randomized-measurement results.
However the latter are consistently much closer to the ideal target value.
This is in agreement with numerical simulations (not shown) that include various noise sources except non-local gate errors.
We find qualitatively similar results for other pairs of orthogonal target states.

\textit{Imperfect transfer}.\textemdash
In addition to local errors on each node, the realization of distributed protocols is limited by erroneous and leaky state transfer over a waveguide, \textit{e.g.}, an optical fibre, a cryogenic microwave link or an acoustic waveguide.
Possible error sources include particle loss and non-linear dispersion of the waveguide.
On-chip and state-transfer errors have to be considered together:
\textit{e.g.}, photons with narrower bandwidth are more robust against distortion and yield higher transfer fidelities \cite{Penas2021};
however, protocols involving narrower photons require more time, which leads to a trade-off between imperfect state transfer and stationary qubit decoherence.
Here we estimate achievable fidelities of the distributed SWAP (S1) and Bell-basis (S2) protocols, based on a simple scaling analysis.

We may distinguish between four main sources of error, \textit{i.e.},
(\textit{i}) imperfect state transfer with fidelities $\mathcal{F}_\mathrm{st} < 1$,
(\textit{ii}) finite qubit decoherence rates $\Gamma = 1/T_1$,
(\textit{iii}) gate fidelities $\mathcal{F}_\mathrm{g} < 1$,
and (\textit{iv}) readout errors $1-\mathcal{F}_\mathrm{r}>0$.
Here $\mathcal{F}_\mathrm{g}$ denotes the joint fidelity of each separate block in Figs.~\ref{fig:distributed_swap_schemes}(a) and (b).
Based on Eqs.~\eqref{eq:operation-cswap} and \eqref{eq:operation-block1} we estimate the \textit{protocol success rates}
\begin{equation}\label{eq:fidelity-swap-estimate}
\begin{aligned}
    p_\mathrm{S1} & = \left [ \mathcal{F}_\mathrm{st} \exp \left ( - \Gamma T_\mathrm{block} \right ) \right ]^{2n} \mathcal{F}_\mathrm{g}^n \mathcal{F}_\mathrm{r}, \\
    p_\mathrm{S2} & = \left [ \mathcal{F}_\mathrm{st} \exp \left ( - \Gamma T_\mathrm{block} \right ) \mathcal{F}_\mathrm{g} \mathcal{F}_\mathrm{r} \right ]^n,
\end{aligned}
\end{equation}
where $T_\mathrm{block}$ denotes the duration of a single block in Fig.~\ref{fig:distributed_swap_schemes}.
The poorer scaling of S1 is related to the fact that the ancilla qubit is measured only at the end of the protocol.
If the qubits are not sent back to the sending node in Fig.~\ref{fig:distributed_swap_schemes}(c), the scaling is improved, \textit{i.e.}, $p_\mathrm{S1} \propto \mathcal{F}_\mathrm{st}^{n}$.

We note that the discussed methods are supposed to be useful for reasonably good devices.
For a fixed error rate, the fidelity of states on each device will itself decrease exponentially with system size and the estimation procedures become less useful.
On the other hand, they may be useful when applied to the logical qubits of fault-tolerant quantum computers.
An analysis of these aspects is beyond the scope of this work.

\begin{figure*}
\centering
\includegraphics[width=0.75\textwidth]
{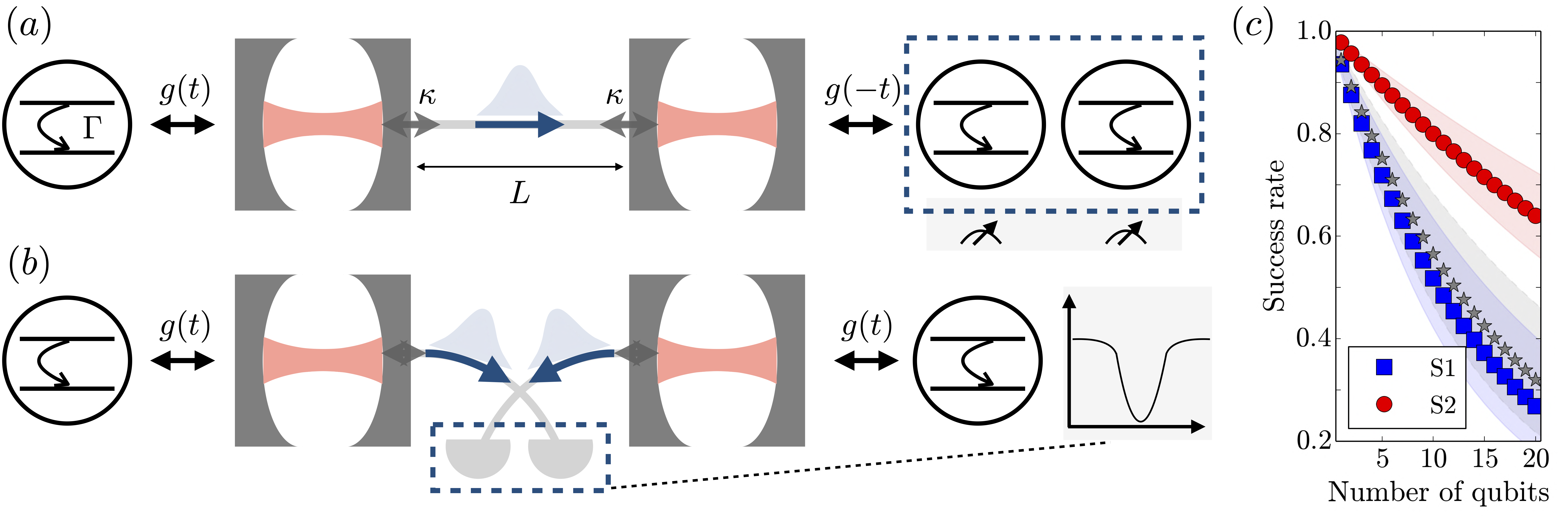}
    \caption{Schematic illustration of quantum optical implementation of state-transfer and overlap-estimation schemes.
    (\textit{a}) Qubits with finite relaxation rate $\Gamma$ interacting with cavity mode via time-dependent coupling $g(t)$.
    Cavity decay rate $\kappa$ determines photon loss into waveguide of length $L$ that is connected to another cavity at the second node of a network.
    High-fidelity state transfer between the nodes may be realized by choosing time-symmetric $g(t)$ and coupling the qubit at the second node to its cavity via the time-reversed coupling $g(-t)$ \cite{Cirac1997}.
    Scheme S2 depicted in Fig.~\ref{fig:distributed_swap_schemes} requires Bell-basis measurements, indicated by dashed box.
    State overlap may be deduced via classical post-processing.
    (\textit{b}) In scheme S4, the state overlap between distant quantum states is estimated from a Hong-Ou-Mandel interference experiment, \textit{e.g.}, using microwave photons.
    The state overlap can be obtained from the coincidence counts on the output detectors and classical post-processing.
    (\textit{c}) Success rate of schemes S1 (blue squares and shaded region) and S2 (red squares and shaded region) as estimated from Eq.~\eqref{eq:fidelity-swap-estimate}.
    Additional data points (gray stars and shaded region) show estimated success rate for S1 without sending back the qubits to the sending node in Fig.~\ref{fig:distributed_swap_schemes}(c).
    Parameters: $1 - \mathcal{F}_\mathrm{st} = (0.75 - 1) \times 10^{-2}$, $\Gamma T_\mathrm{block} = (2.5 - 7.5) \times 10^{-3}$, $1 - \mathcal{F}_\mathrm{r} = 10^{-3}$, $1 - \mathcal{F}_\mathrm{g} = (2.5 - 5.0) \times 10^{-2}$ (S1) and $1 - \mathcal{F}_\mathrm{g} = (0.5 - 1) \times 10^{-2}$ (S2), respectively.
    }
    \label{fig:implementations}
\end{figure*}

\section{Implementations}\label{sec:implementations}

The protocols discussed in Sec.~\ref{sec:cross-platform-verification} may be implemented with several physical platforms.
The main ingredient is quantum communication between spatially separated nodes \cite{Cirac1997}.
Motivated by recent progress, we discuss prospects for future experimental investigations with different physical systems, and provide realistic estimates for achievable fidelities and system sizes in state-of-the-art architectures.

\subsection{Superconducting circuits}

Most physical building blocks for implementing the schemes discussed in Sec.~\ref{ssec:sequential-tests} have already been demonstrated in experiments with SC circuits.
All protocols rely on quantum transmission across a quantum link, but they differ in terms of various details.
Scheme S1 is based on the standard SWAP test and has a higher gate count than the remaining schemes.
The circuit of the SWAP test involves a controlled-SWAP (Fredkin) gate, which may be decomposed into two controlled-NOT gates and one Toffoli gate, see Fig.~\ref{fig:distributed_swap_schemes}(a).
The latter has been realized with SC qubits already in \cite{Fedorov2013}, and state-of-the-art architectures can realize high-fidelity three-qubit gates \cite{Kim2022}.
Scheme S2 requires only $n$ two-qubit gates, which are routinely realized with high fidelity in current setups.
The realization of scheme S3 relies on interactions between traveling and stationary qubits \cite{Duan2004}, which have recently been harnessed to implement a qubit-photon controlled-phase gate \cite{Reuer2022}.
The implementation of scheme S4 requires two-photon interference, and Hong-Ou-Mandel coalescence at microwave frequencies has been demonstrated before \cite{Lang2013}.
Here, the coincidence count rate of photonic states emitted from separate nodes, cf.~Fig.~\ref{fig:implementations}(b), depends on the details of their mode functions.
With rapid improvements in chip and transmission quality, the realization of cross-verification schemes may soon be possible with SC circuits.

\textit{Fidelities}.\textemdash
SC qubits feature relatively fast two-qubit gate times $\sim (10-100)~$ns with high fidelities $>0.99$ and qubit coherence times exceeding $100~\mu$s \cite{Kjaergaard2020}.
Moreover, quantum processors based on superconducting qubits can be connected by coherent quantum links in state-of-the-art experiments, and they are well-suited for network-based quantum operations as required for implementing distributed algorithms.
A possible physical realization of the basic building block, \textit{i.e.}, quantum state transfer, is schematically depicted in Fig.~\ref{fig:implementations}(a) and discussed in detail in Ref.~\cite{Kurpiers2018}.
It involves a waveguide of length $L$ that connects two resonators, each coupled to an artificial atom via a time-dependent coupling $g(t)$.
The coupling between cavity and waveguide is governed by the cavity decay rate $\kappa$.
The qubit-resonator coupling $g(t)$ can be chosen such that the state of the qubit is mapped onto a time-symmetric photonic state which travels through the waveguide and is reabsorbed at the second node, realizing high-fidelity state transfer only limited by physical information loss.
Recently this protocol has successfully been implemented to connect two transmon qubits located in two spatially separated dilution refrigerators, achieving an average state-transfer fidelity of $\mathcal{F}_\mathrm{st} \approx 0.86$ \cite{Magnard2020}.
Future improvements are expected to lead to much higher values \cite{Penas2021}.
Based on realistic and optimistic coherence times, gate and transfer fidelities, in Fig.~\ref{fig:implementations}(c) we estimate achievable protocol success probabilities according to Eq.~\eqref{eq:fidelity-swap-estimate}.
We project that schemes based on Bell-basis measurements may allow for success rates $> 0.7$ at system sizes of up to $n \approx 20$ qubits per node.

\subsection{Trapped ions}
Owing to their long trapping lifetimes and coherence times, trapped ions are a main contender for realizing useful quantum algorithms and communication protocols.
Deterministic ion-photon interfaces based on optical cavities enable quantum state transfer between different nodes of a network via the polarization degree of freedom of photons \cite{Northup2014}.
Such photonic interconnects may be used to couple separate ion registers \cite{Duan2010,Monroe2014} and entangle distant ions \cite{Moehring2007}.
Recently, remote entanglement of trapped-ion qubits has been achieved at rates approaching those of local operations, and heralded Bell pairs have been created with high reported fidelities ($\approx 0.94$) and at fast rates \cite{Stephenson2020}.
Anticipating further progress in this field, soon trapped-ion systems may thus be used to realize proofs of concept of entanglement-based verification protocols such as those mentioned in Sec.~\ref{ssec:compare-computations} or in the context of interactive proofs \cite{Gheorgiu2019}.
Improved photon-based information transfer may enable the implementation of scalable, distributed SWAP tests with trapped-ion platforms.
Several important ingredients have already been demonstrated in the past.
A three-qubit Toffoli gate with a fidelity $> 0.7$ has been implemented in Ref.~\cite{Monz2008}.
Hong-Ou-Mandel interference experiments may be performed on photons and phonons \cite{Toyoda2015}.
Single photons emitted by trapped ions, separated by $\sim 2~$m, show highly suppressed coincidence count rates for indistinguishable photons as compared to distinguishable photons \cite{Kim2021}.
Future improvements of photonic mode matching and back-reflection will further improve these results.

\subsection{Other platforms}

There is a variety of other physical systems that are being actively explored as constituents of quantum networks, including other information carriers than photons.
Based on significant improvements in various fields, it will be interesting to further investigate their case-specific advantages for network-based protocols.
For instance, some setups may be ideally suited for implementing high-fidelity cross-platform verification protocols based on interference experiments while others are ideally suited for gate-based schemes.
Such unique advantages of specific platforms can also be exploited by interlinking different platforms, as has been achieved, \textit{e.g.}, for a heterogenous network of a cold atomic ensemble optically connected to a rare-earth-doped crystal \cite{Maring2017}.
Solid-state architectures may be coherently connected by phononic quantum links, \textit{e.g.}, based on surface acoustic waves as proposed \cite{Schuetz2015} and realized \cite{Bienfait2019} in recent years.
Demonstrations of qubit teleportation between distant color defects in diamond pave the way for increased communication rates and fidelities in quantum networks connecting separate solid-state spin qubit registers \cite{Hermans2022}.
Quantum-dot single-photon sources can produce highly indistinguishable photons that may be useful for various communication protocols \cite{Lu2021}, including scheme S4 outlined in Sec.~\ref{ssec:sequential-tests}.
Another possibility are tweezer arrays of Rydberg atoms \cite{Saffman2010}, in which high fidelity gates in large arrays of qubits have been realized \cite{Levine2019}.
More challenging in this platform are quantum links, but substantial progress has been made for example with nanophotonic interfaces \cite{Dordevic2021}.
Proof-of-principle experiments could also be done by dividing a single tweezer array into two spatially separated arrays, between which atoms can be transported coherently \cite{Bluvstein2022}.
All these and a range of other developments may enable near-term realizations of verification protocols in small- and intermediate-scale quantum networks, that may be tested and eventually deployed in large-scale networks \cite{Wehner2018}.

\section{Conclusions \& outlook \label{sec:outlook}}

In conclusion, we have described verification techniques for quantum states and computations in quantum networks.
From a practical perspective, near-term realization of efficient verification protocols will provide invaluable benchmarking tools and essential means for gaining trust and confidence in quantum devices, and their ability to perform computations or prepare specific quantum states.
On a more fundamental level, access to coherent quantum state transfer and SWAP tests provides an exponential advantage over protocols that rely on classical communication, in terms of experimentally required resources \cite{Aharonov2022}, at least if one discounts errors in the protocols.
Future studies are required for a detailed analysis of erorr sources, and how to correct for them.

Motivated by recent experimental achievements, which have set the stage for high-fidelity state transfer between distant quantum processors, we have analyzed prospects for cross-platform verification.
To this aim, we have distinguished between four different state-comparison schemes based on the SWAP test and Bell-basis measurements, and discussed their scaling of required resources with system size.
While the communication complexity scales linearly with input size in all considered cases, Bell-basis measurements are generally less demanding and may already be realized with state-of-the-art experimental setups.
We estimate that this may allow for relatively faithful state comparisons of $n \approx (10-20)$-qubit states in forthcoming superconducting-circuit architectures.
Protocols for state comparison also provide means for efficiently comparing quantum computations.
We have considered three ways for how this connection can be made explicit, using the channel-state duality, averaging over random initial states and entanglement witnesses.

The dawn of quantum networks connecting intermediate-scale quantum processors at each node provides an ideal test bed for a variety of other verification paradigms.
For instance, variations of the distributed SWAP test may be utilized to distinguish states belonging to different entanglement classes \cite{Foulds2021}.
In quantum networks with multiple nodes, distributed protocols may be useful for multivariate trace estimation \cite{Quek2022}.
Other exciting directions to be explored in near-term quantum networks include the experimental exploration of quantum algorithmic measurements \cite{Aharonov2022}, error mitigation \cite{Koczor2021}, multi-core computing \cite{Jnane2022} and interactive verification protocols \cite{Gheorgiu2019}.
While many of these tasks are still challenging for implementations, some have milder resource requirements and may soon be implemented in practice.
Deploying such protocols will help to establish a broad experimental toolbox for quantum verification.

\section*{Acknowledgements}
We thank Jordi Tura, Benjamin Schiffer, Alexandru Gheorghiu and Fran\c {c}ois Bienvenu for helpful discussions.
J.~K. gratefully acknowledges support from Dr. Max R\"ossler, the Walter Haefner Foundation and the ETH Z\"urich Foundation.
J.~K. and J.~I.~C. gratefully acknowledge support from the European Union’s Horizon 2020 FET-Open project SuperQuLAN (899354) and the Deutsche Forschungsgemeinschaft (DFG, German Research Foundation) under Germany’s Excellence Strategy – EXC-2111 – 390814868.
We acknowledge the use of IBM Quantum services for this work.
The views expressed are those of the authors, and do not reflect the official policy or position of IBM or the IBM Quantum team.
In this paper we used the five-qubit devices $\mathrm{ibmq}\textunderscore\mathrm{manila}$ and $\mathrm{ibm}\textunderscore\mathrm{perth}$, as well as the $16$-qubit machine $\mathrm{ibmq}\textunderscore\mathrm{guadalupe}$, which are IBM Quantum Falcon Processors.
This research is part of the Munich Quantum Valley, which is supported by the Bavarian state government with funds from the Hightech Agenda Bayern Plus.

\bibliography{main}

\clearpage
\onecolumngrid

\appendix

\begin{center}
\textbf{\large Supplemental Materials: Cross-Platform Verification in Quantum Networks}
\end{center}

This Supplemental Material is structured as follows.
In Sec.~\ref{app:state-verification} we provide a brief summary of the SWAP test and Bell-basis measurements and show experimental results of both tests, obtained on a superducting quantum device, applied to a simple example.
Sec.~\ref{app:sequential-tests} summarizes the basic steps of the distributed protocols from the main text.
In Sec.~\ref{app:confidence} we summarize statistical tools that can be used to derive confidence intervals of the fidelity estimates.
Sec.~\ref{app:comp-state-three-methods} contains a summary of the methods introduced in Sec.~\ref{ssec:compare-computations}, targeted at comparing quantum operations at different nodes.

\section{Comparison of quantum states \label{app:state-verification}}

Full quantum state tomography can be considered the gold standard for verification and benchmarking of small quantum systems \cite{Cramer2010}.
QST yields access to all entries of the density matrix of a quantum state, but this level of information comes at a highly increased cost of required measurement resources.
Despite the existence of improved tomography protocols based on compressed sensing \cite{Gross2010}, the required measurement budget still scales exponentially with the system size.
Luckily, in many applications a full reconstruction of the density matrix is not needed.
For example, sometimes a given state shall be compared to an ideal target state, or compared to another unknown state.

\textit{Fidelity estimation}.\textemdash
Two quantum states $\rho$ and $\tilde \rho$ may be compared by measuring their Uhlmann fidelity $\mathcal{F}(\rho,\tilde \rho) = \left ( \mathrm{tr}\sqrt{\rho\sqrt{\tilde \rho}\rho}\right )^2$, or by the expression \eqref{eq:fidelity} introduced in the main text.
In case one state is pure, $\mathcal{F}(\rho,\tilde \rho) = \mathrm{tr}\left ( \rho\tilde \rho \right)$ and if both $\rho = |\Psi\rangle\langle\Psi|$ and $\tilde \rho = |\tilde \Psi\rangle\langle\tilde \Psi|$ are pure, $\mathcal{F}(\ket{\Psi},\ket{\tilde \Psi}) = |\langle \Psi | \tilde \Psi \rangle|^2$ is simply the modulus squared of the overlap between the state vectors.
In many scenarios, a quantum state shall be compared to an ideal target state to assess the quality of an experiment and its computational output.
In that case, direct fidelity estimation provides a means to estimate the fidelity with respect to the target state using only Pauli measurements, and fewer measurements than QST \cite{Flammia2011}.
For special classes of states, which are of importance in quantum information processing, there exist methods to efficiently estimate the fidelity with only polynomial measurement costs.
For example, Dicke states \cite{Liu2019}, hypergraph states \cite{Zhu2019} and GHZ states \cite{Zihao2020} allow for efficient fidelity esimation techniques using only local measurements.
In other settings, the task might be to estimate the fidelity of two \textit{a priori} unknown quantum states.
In that case, $F(\rho,\tilde \rho)$ may not be inferred from prior knowledge of a given target state.

\begin{figure}[b]
\centering
\includegraphics[width=0.48\textwidth]{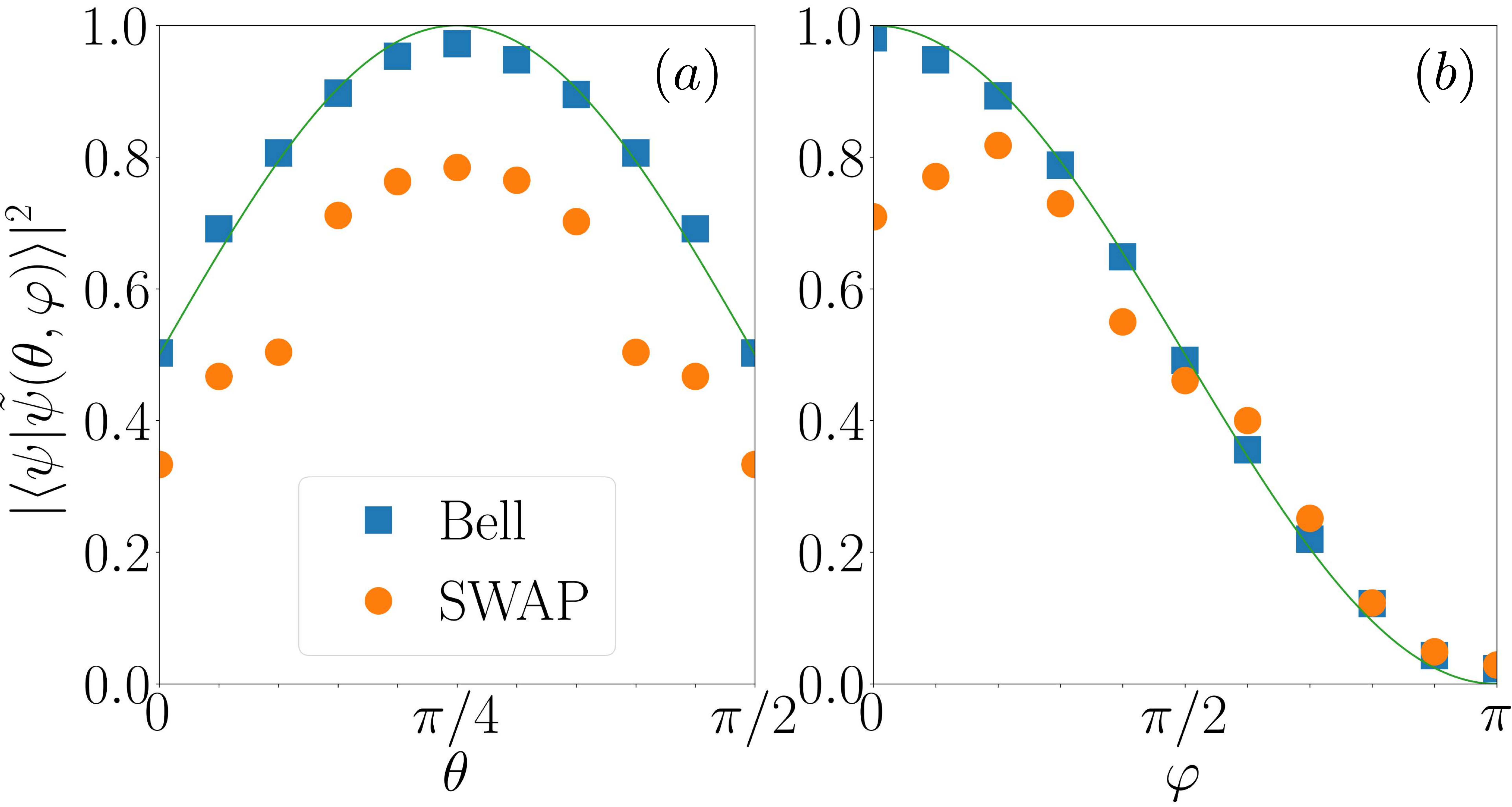}
    \caption{Fidelity $\mathcal{F}(|\psi\rangle, |\tilde \psi\rangle)$ for states $|\psi\rangle = (|0\rangle + |1\rangle)/\sqrt{2}$ and $|\tilde \psi (\theta, \varphi) \rangle = (\cos(\theta) |0\rangle + \sin(\theta) e^{i\varphi}|1\rangle)$ prepared on IBM quantum processor $\mathrm{ibm}\textunderscore\mathrm{manila}$, obtained via measurements in computational basis and classical post-processing and
    (\textit{a}) as a function of $\theta$ at $\varphi = 0$,
    (\textit{b}) as a function of $\varphi$ at $\theta = \pi/4$.
    Green line is exact overlap for ideal target states.
    Each data point is constructed from $65536$ shots.
    Statistical error bars are smaller than labels.
    }
    \label{fig:fidelity}
\end{figure}

\textit{SWAP test}.\textemdash
The standard procedure for obtaining the state overlap between two unknown states $\rho$ and $\tilde \rho$ is based on the so-called SWAP test \cite{Buhrman2001}.
For simplicity, let us consider two pure quantum states $\ket{\Psi}$ and $\ket{\tilde \Psi}$ and an ancillary qubit denoted by $\ket{\cdot}_\mathrm{anc}$.
The quantum circuit depicted in Fig.~\ref{fig:distributed_swap_schemes}(a) describes the SWAP test to estimate the modulus of the state overlap, $|\langle\Psi|\tilde \Psi\rangle|$.
After execution of the circuit and before the measurement, the entire system is in the state
\begin{eqnarray}\label{eq:psi-final-swap}
    \ket{\Psi}_\mathrm{tot} = \frac{1}{2} & \ket{0}_\mathrm{anc} & \otimes \left ( \ket{\Psi} \ket{\tilde \Psi} +  \ket{\tilde \Psi} \ket{\Psi} \right ) + \nonumber \\
    \frac{1}{2} & \ket{1}_\mathrm{anc} & \otimes \left ( \ket{\Psi} \ket{\tilde \Psi} -  \ket{\tilde \Psi} \ket{\Psi} \right ),
\end{eqnarray}
and the outcome of a computational-basis measurement on the ancilla qubit yields $1$ with probability $(1-|\braket{\Psi|\tilde \Psi}|^2)/2$.
Repeated execution of the SWAP test may be used to deduce $|\langle \Psi | \tilde \Psi \rangle|$.
The only measurement in the SWAP test is performed on the ancilla, not on $\ket{\Psi}$ or $\ket{\tilde \Psi}$, which makes the test \textit{non-destructive}.

\textit{Bell-basis measurements}.\textemdash
Bell-basis measurements, as described by the circuit in Fig.~\ref{fig:distributed_swap_schemes}(b), may be regarded as an alternative to the SWAP test that requires no ancilla and less intensive use of rather costly two-qubit gates, at the expense of destructive state measurements and additional classical post-processing.
This has been previously described in Refs.~\cite{Escartin2013,Cincio_2018}.
Measurements on both subsystems yield two bitstrings $B_1 ... B_n$ and $C_1 ... C_n$, cf.~Fig.~\ref{fig:distributed_swap_schemes}(b).
The parity $P$ of the bitwise AND operation, $P_i = B_i \land C_i$ ($i = 1, ..., n$), determines whether $\ket{\Psi}$ and $\ket{\tilde \Psi}$ pass the test.
If $P = \sum_i P_i$ is even, they pass and otherwise fail the test.
Repeated execution yields an estimate of the state overlap, similar as described above.
Note that the circuit has a constant depth of two, assuming full connectivity between the qubit pairs in Fig.~\ref{fig:distributed_swap_schemes}(b), which compares favorably with the linearly growing circuit depth in Fig.~\ref{fig:distributed_swap_schemes}(a).
This is of practical relevance, as discussed in Ref.~\cite{Cincio_2018} and highlighted in Fig.~\ref{fig:fidelity}.
Here, we compare the analytically known fidelity $\mathcal{F}(\ket{\psi}, \ket{\tilde \psi(\theta, \varphi)})$, for two given states as a function of parameters $\theta$ and $\varphi$, with an estimate obtained from executing the circuits in Fig.~\ref{fig:distributed_swap_schemes} on a superconducting quantum processor.
The ancilla-free algorithm performs significantly better due to its reduced gate count and shorter circuit.

\section{Distributed overlap estimation \label{app:sequential-tests}}

In this appendix, we summarize the steps of the distributed protocols discussed in Sec.~\ref{sec:cross-platform-verification}.

\subsection{Scheme S1}

The scheme S1 is depicted in Fig.~\ref{fig:distributed_swap_schemes}(c).
Initially $\ket{\Psi}_\mathrm{n1}$ and $\ket{\tilde \Psi}_\mathrm{n2}$ are prepared at two separate nodes of a quantum network.
The storage and operational ancilla qubits $\ket{0}_\mathrm{st}$ and $\ket{0}_\mathrm{op}$ are located at the second node.
Using a state-transfer protocol $\mathcal{T}$, the full state is mapped onto
\begin{eqnarray}
     & & \underset{\mathrm{node} \ 1}{\underline{\ket{\Psi}_ \mathrm{n1}}} \otimes  \underset{\mathrm{node} \ 2}{\underline{\ket{0}_\mathrm{st} \otimes \ket{\tilde \Psi}_\mathrm{n2} \otimes \ket{0}_\mathrm{op}}} \nonumber \\
    & \overset{\mathcal{T}}{\longrightarrow} &
    \left ( \sum_{i_2, ...} a_{i_2, ...}\ket{0,i_2,...} \otimes \alpha_1 \ket{0}_\mathrm{st} + \sum_{i_2, ...} b_{i_2, ...}\ket{0,i_2,...} \otimes \beta_1 \ket{1}_\mathrm{st} \right ) \otimes \ket{\tilde\Psi} \otimes \ket{0}_\mathrm{op} \nonumber \\
    & = &
    \left ( \sum_{i_2, ...} a_{i_2, ...}\ket{0,i_2,...} \otimes \alpha_1 \ket{0}_\mathrm{st} + \sum_{i_2, ...} b_{i_2, ...}\ket{0,i_2,...} \otimes \beta_1 \ket{1}_\mathrm{st} \right ) \nonumber \\
    & & \otimes \left [ \tilde \alpha_1 \ket{0}_{\tilde 1} \otimes \sum_{i_2, ...} \tilde a_{i_2, ...} \ket{i_2, ...} + \tilde \beta_1 \ket{1}_{\tilde 1} \otimes \sum_{j_2, ...} \tilde b_{j_2, ...} \ket{j_2, ...} \right ] \otimes \ket{0}_\mathrm{op} \nonumber
\end{eqnarray}    
The operational ancilla takes the role of the auxiliary qubit in Fig.~\ref{fig:distributed_swap_schemes}(a).
A Hadamard gate is applied to the operational ancilla, to perform a SWAP test between the two nodes:    
\begin{eqnarray}    
    & \overset{\mathrm{H}}{\longrightarrow} &
    \left ( \sum_{i_2, ...} a_{i_2, ...}\ket{0,i_2,...} \otimes \alpha_1 \ket{0}_\mathrm{st} + \sum_{i_2, ...} b_{i_2, ...}\ket{0,i_2,...} \otimes \beta_1 \ket{1}_\mathrm{st} \right ) \nonumber \\
    & & \otimes \left [ \tilde \alpha_1 \ket{0}_{\tilde 1} \otimes \sum_{i_2, ...} \tilde a_{i_2, ...} \ket{i_2, ...} + \tilde \beta_1 \ket{1}_{\tilde 1} \otimes \sum_{j_2, ...} \tilde b_{j_2, ...} \ket{j_2, ...} \right ] \otimes \frac{\ket{0}_\mathrm{op}+\ket{1}_\mathrm{op}}{\sqrt{2}}, \nonumber
\end{eqnarray}
followed by a Fredkin (controlled-SWAP) gate,
\begin{eqnarray}
    & \overset{\mathrm{cSWAP}}{\longrightarrow} & \frac{1}{\sqrt{2}} \times \Bigg [
    \left ( \sum_{i_2, ...} a_{i_2, ...}\ket{0, i_2,...}_{\mathrm{n}1} \right ) \otimes
    \bigg ( \alpha_1 \tilde \alpha_1 \ket{0}_\mathrm{st} \otimes \ket{0}_\mathrm{\tilde 1} \bigg )
    \otimes \left ( \sum_{j_2, ...} \tilde a_{i_2, ...}\ket{j_2,...}_{\mathrm{n}2} \right ) \otimes \ket{0}_\mathrm{op} \nonumber \\
    & & \ \ \ \ \ \ \ \ \ \ \ +
    \left ( \sum_{i_2, ...} b_{i_2, ...}\ket{0, i_2,...}_{\mathrm{n}1} \right ) \otimes \bigg ( \beta_1 \tilde \alpha_1 \ket{1}_\mathrm{st} \otimes \ket{0}_\mathrm{\tilde 1}   \bigg ) \otimes \left ( \sum_{j_2, ...} \tilde a_{i_2, ...}\ket{j_2,...}_{\mathrm{n}2} \right ) \otimes \ket{0}_\mathrm{op} \nonumber \\
    & & \ \ \ \ \ \ \ \ \ \ \ +
    \left ( \sum_{i_2, ...} a_{i_2, ...}\ket{0, i_2,...}_{\mathrm{n}1} \right ) \otimes   
    \bigg ( \alpha_1 \tilde \beta_1 \ket{0}_\mathrm{st} \otimes \ket{1}_\mathrm{\tilde 1} \bigg )
    \otimes \left ( \sum_{j_2, ...} \tilde b_{i_2, ...}\ket{j_2,...}_{\mathrm{n}2} \right ) \otimes \ket{0}_\mathrm{op} \nonumber \\
    & & \ \ \ \ \ \ \ \ \ \ \ +
    \left ( \sum_{i_2, ...} b_{i_2, ...}\ket{0, i_2,...}_{\mathrm{n}1} \right ) \otimes 
    \bigg ( \beta_1 \tilde \beta_1 \ket{1}_\mathrm{st} \otimes \ket{1}_\mathrm{\tilde 1} \bigg )  
    \otimes \left ( \sum_{j_2, ...} \tilde b_{i_2, ...}\ket{j_2,...}_{\mathrm{n}2} \right ) \otimes \ket{0}_\mathrm{op} \nonumber \\
    & & \ \ \ \ \ \ \ \ \ \ \ + 
    \left ( \sum_{i_2, ...} a_{i_2, ...}\ket{0, i_2,...}_{\mathrm{n}1} \right ) \otimes
    \bigg ( \alpha_1 \tilde \alpha_1 \ket{0}_\mathrm{st} \otimes \ket{0}_\mathrm{\tilde 1} \bigg )
    \otimes \left ( \sum_{j_2, ...} \tilde a_{i_2, ...}\ket{j_2,...}_{\mathrm{n}2} \right ) \otimes \ket{1}_\mathrm{op} \nonumber \\
    & & \ \ \ \ \ \ \ \ \ \ \ +
    \left ( \sum_{i_2, ...} b_{i_2, ...}\ket{0, i_2,...}_{\mathrm{n}1} \right ) \otimes \bigg ( \tilde \beta_1 \alpha_1 \ket{1}_\mathrm{st} \otimes \ket{0}_\mathrm{\tilde 1}   \bigg ) \otimes \left ( \sum_{j_2, ...} \tilde a_{i_2, ...}\ket{j_2,...}_{\mathrm{n}2} \right ) \otimes \ket{1}_\mathrm{op} \nonumber \\
    & & \ \ \ \ \ \ \ \ \ \ \ +
    \left ( \sum_{i_2, ...} a_{i_2, ...}\ket{0, i_2,...}_{\mathrm{n}1} \right ) \otimes
    \bigg ( \tilde \alpha_1 \beta_1 \ket{0}_\mathrm{st} \otimes \ket{1}_\mathrm{\tilde 1} \bigg )  
    \otimes \left ( \sum_{j_2, ...} \tilde b_{i_2, ...}\ket{j_2,...}_{\mathrm{n}2} \right ) \otimes \ket{1}_\mathrm{op} \nonumber \\
    & & \ \ \ \ \ \ \ \ \ \ \ +
    \left ( \sum_{i_2, ...} b_{i_2, ...}\ket{0, i_2,...}_{\mathrm{n}1} \right ) \otimes
    \bigg ( \beta_1 \tilde \beta_1 \ket{1}_\mathrm{st} \otimes \ket{1}_\mathrm{\tilde 1} \bigg )
    \otimes \left ( \sum_{j_2, ...} \tilde b_{i_2, ...}\ket{j_2,...}_{\mathrm{n}2} \right ) \otimes \ket{1}_\mathrm{op} \Bigg ] \nonumber \\
\end{eqnarray}

Now the qubit is sent back to its original location at node 1.
Then the state reads

\begin{eqnarray}
   & \overset{\mathcal{T}^\dagger}{\longrightarrow} & \frac{1}{\sqrt{2}} \times \Bigg [
    \left ( \sum_{i_1, ...} c_{i_1, ...}\ket{i_1, ,...}_{\mathrm{n}1} \right ) \otimes
    \left ( \sum_{j_1, ...} \tilde c_{i_1, ...}\ket{j_1,...}_{\mathrm{n}2} \right ) \otimes \ket{0}_\mathrm{st} \otimes \ket{0}_\mathrm{op} \nonumber \\
    & & \ \ \ \ \ \ \ \ \ \ \ +
    \left ( \sum_{i_2, ...} a_{i_2, ...}\ket{0, i_2,...}_{\mathrm{n}1} \right ) \otimes
    \bigg ( \alpha_1 \tilde \alpha_1 \ket{0}_1 \otimes \ket{0}_\mathrm{\tilde 1} \bigg )
    \otimes \left ( \sum_{j_2, ...} \tilde a_{i_2, ...}\ket{j_2,...}_{\mathrm{n}2} \right ) \otimes \ket{0}_\mathrm{st} \otimes \ket{1}_\mathrm{op} \nonumber \\
    & & \ \ \ \ \ \ \ \ \ \ \ +
    \left ( \sum_{i_2, ...} b_{i_2, ...}\ket{0, i_2,...}_{\mathrm{n}1} \right ) \otimes \bigg ( \tilde \beta_1 \alpha_1 \ket{1}_1 \otimes \ket{0}_\mathrm{\tilde 1}   \bigg ) \otimes \left ( \sum_{j_2, ...} \tilde a_{i_2, ...}\ket{j_2,...}_{\mathrm{n}2} \right ) \otimes \ket{0}_\mathrm{st} \otimes \ket{1}_\mathrm{op} \nonumber \\
    & & \ \ \ \ \ \ \ \ \ \ \ +
    \left ( \sum_{i_2, ...} a_{i_2, ...}\ket{0, i_2,...}_{\mathrm{n}1} \right ) \otimes  
    \bigg ( \tilde \alpha_1 \beta_1 \ket{0}_1 \otimes \ket{1}_\mathrm{\tilde 1} \bigg )  
    \otimes \left ( \sum_{j_2, ...} \tilde b_{i_2, ...}\ket{j_2,...}_{\mathrm{n}2} \right ) \otimes \ket{0}_\mathrm{st} \otimes \ket{1}_\mathrm{op} \nonumber \\
    & & \ \ \ \ \ \ \ \ \ \ \ +
    \left ( \sum_{i_2, ...} b_{i_2, ...}\ket{0, i_2,...}_{\mathrm{n}1} \right ) \otimes  
    \bigg ( \beta_1 \tilde \beta_1 \ket{1}_1 \otimes \ket{1}_\mathrm{\tilde 1} \bigg )
    \otimes \left ( \sum_{j_2, ...} \tilde b_{i_2, ...}\ket{j_2,...}_{\mathrm{n}2} \right ) \otimes \ket{0}_\mathrm{st} \otimes \ket{1}_\mathrm{op} \Bigg ]. \nonumber \\
\end{eqnarray}
This is the intermediate state of the full system at the end of block 1, cf.~Fig.~\ref{fig:distributed_swap_schemes}(c).
This procedure is repeated for all $n$ pairs of qubits, before the final Hadamard gate in Fig.~\ref{fig:distributed_swap_schemes}(a) maps $\ket{0}_\mathrm{op} \rightarrow (\ket{0}_\mathrm{op} + \ket{1}_\mathrm{op})/\sqrt{2}$ and $\ket{1}_\mathrm{op} \rightarrow (\ket{0}_\mathrm{op} - \ket{1}_\mathrm{op})/\sqrt{2}$.
If both states $\ket{\Psi}_\mathrm{n1}$ and $\ket{\tilde \Psi}_\mathrm{n2}$ were single-qubit states, the protocol would end here with a measurement of the operational ancilla in the computational basis.
Otherwise, the test continues in full analogy to the ordinary SWAP test on a single device, cf.~Fig.~\ref{fig:distributed_swap_schemes}(a).
The final state before the measurement will be given by Eq.~\eqref{eq:psi-final-swap}.
If all coefficients $c_{\vec{i}}$ and $\tilde c_{\vec{i}}$ in Eq.~\eqref{eq:states-1} coincide, the ancilla will always be found in the state $\ket{0}_\mathrm{op}$.

\subsection{Scheme S2}

Similarly, the step-by-step evolution of the full state in scheme S2 starts with state transfer from node 1 to node 2, followed by a $\mathrm{CNOT}$ gate at node 2,
\begin{eqnarray}
     & & \underset{\mathrm{node} \ 1}{\underline{\ket{\Psi}_\mathrm{n1}}} \otimes  \underset{\mathrm{node} \ 2}{\underline{\ket{0}_\mathrm{st} \otimes \ket{\tilde \Psi}_\mathrm{n2}}} \nonumber \\
    & \overset{\mathcal{T}}{\longrightarrow} &
    \left ( \sum_{i_2, ...} a_{i_2, ...}\ket{0,i_2,...}_\mathrm{n1} \otimes \alpha_1 \ket{0}_\mathrm{st} + \sum_{i_2, ...} b_{i_2, ...}\ket{0,i_2,...}_\mathrm{n1} \otimes \beta_1 \ket{1}_\mathrm{st} \right ) \otimes \ket{\tilde\Psi}_\mathrm{n2} \nonumber \\
    & = &     \left ( \sum_{i_2, ...} a_{i_2, ...}\ket{0,i_2,...}_\mathrm{n1} \otimes \alpha_1 \ket{0}_\mathrm{st} + \sum_{i_2, ...} b_{i_2, ...}\ket{0,i_2,...}_\mathrm{n1} \otimes \beta_1 \ket{1}_\mathrm{st} \right ) \nonumber \\
    & & \otimes
    \left [ \tilde \alpha_1 \ket{0}_{\tilde 1} \otimes \sum_{i_2, ...} \tilde a_{i_2, ...} \ket{i_2, ...} + \tilde \beta_1 \ket{1}_{\tilde 1} \otimes \sum_{j_2, ...} \tilde b_{j_2, ...} \ket{j_2, ...} \right ]
    \nonumber \\
    & \overset{\mathrm{cNOT}}{\longrightarrow} & \ \ \ \left ( \sum_{i_2, ...} a_{i_2, ...}\ket{0,i_2,...}_\mathrm{n1} \right ) \otimes \Bigg ( \alpha_1 \tilde \alpha_1 \ket{0}_\mathrm{st} \otimes \ket{0}_{\tilde 1} \Bigg ) \otimes \left ( \sum_{j_2, ...} \tilde a_{j_2, ...}\ket{j_2,...}_\mathrm{n2} \right ) \nonumber \\
    & & +
    \left ( \sum_{i_2, ...} a_{i_2, ...}\ket{0,i_2,...}_\mathrm{n1} \right ) \otimes \Bigg ( \alpha_1 \tilde \beta_1 \ket{0}_\mathrm{st} \otimes \ket{1}_{\tilde 1} \Bigg ) \otimes \left ( \sum_{j_2, ...} \tilde b_{j_2, ...}\ket{j_2,...}_\mathrm{n2} \right ) \nonumber \\
    & & +
    \left ( \sum_{i_2, ...} b_{i_2, ...}\ket{0,i_2,...}_\mathrm{n1} \right ) \otimes \Bigg ( \tilde \alpha_1 \beta_1 \ket{1}_\mathrm{st} \otimes \ket{1}_{\tilde 1} \Bigg ) \otimes \left ( \sum_{j_2, ...} \tilde a_{j_2, ...}\ket{j_2,...}_\mathrm{n2} \right ) \nonumber \\
    & & +
    \left ( \sum_{i_2, ...} b_{i_2, ...}\ket{0,i_2,...}_\mathrm{n1} \right ) \otimes \Bigg ( \tilde \beta_1 \beta_1 \ket{1}_\mathrm{st} \otimes \ket{0}_{\tilde 1} \Bigg ) \otimes \left ( \sum_{j_2, ...} \tilde b_{j_2, ...}\ket{j_2,...}_\mathrm{n2} \right ), \nonumber
\end{eqnarray}
where $\ket{\cdot}_\mathrm{st}$ denotes the auxiliary storage qubit at node 2.
Subsequently a Hadamard gate is applied to the storage qubit which maps the state onto
\begin{eqnarray}
    & \overset{\mathrm{H}}{\longrightarrow} & \frac{1}{\sqrt{2}} \Bigg [ \left ( \sum_{i_2, ...} a_{i_2, ...}\ket{0,i_2,...}_\mathrm{n1} \right ) \otimes \alpha_1 \tilde \alpha_1 \Bigg ( \ket{0}_\mathrm{st} \otimes \ket{0}_{\tilde 1} + \ket{1}_\mathrm{st} \otimes \ket{0}_{\tilde 1} \Bigg ) \otimes \left ( \sum_{j_2, ...} \tilde a_{j_2, ...}\ket{j_2,...}_\mathrm{n2} \right ) \nonumber \\
    & & \ \ \ \ \ \ + \left ( \sum_{i_2, ...} a_{i_2, ...}\ket{0,i_2,...}_\mathrm{n1} \right ) \otimes \alpha_1 \tilde \beta_1 \Bigg ( \ket{0}_\mathrm{st} \otimes \ket{1}_{\tilde 1} + \ket{1}_\mathrm{st} \otimes \ket{1}_{\tilde 1} \Bigg ) \otimes \left ( \sum_{j_2, ...} \tilde b_{j_2, ...}\ket{j_2,...}_\mathrm{n2} \right ) \nonumber \\
    & & \ \ \ \ \ \ + \left ( \sum_{i_2, ...} b_{i_2, ...}\ket{0,i_2,...}_\mathrm{n1} \right ) \otimes \tilde \alpha_1 \beta_1 \Bigg ( \ket{0}_\mathrm{st} \otimes \ket{1}_{\tilde 1} - \ket{1}_\mathrm{st} \otimes \ket{1}_{\tilde 1} \Bigg ) \otimes \left ( \sum_{j_2, ...} \tilde a_{j_2, ...}\ket{j_2,...}_\mathrm{n2} \right ) \nonumber \\
    & & \ \ \ \ \ \ + \left ( \sum_{i_2, ...} b_{i_2, ...}\ket{0,i_2,...}_\mathrm{n1} \right ) \otimes \beta_1 \tilde \beta_1 \Bigg ( \ket{0}_\mathrm{st} \otimes \ket{0}_{\tilde 1} - \ket{1}_\mathrm{st} \otimes \ket{0}_{\tilde 1} \Bigg ) \otimes \left ( \sum_{j_2, ...} \tilde b_{j_2, ...}\ket{j_2,...}_\mathrm{n2} \right ) \nonumber \\
\end{eqnarray}
For single-qubit states with $\alpha_1 = \tilde \alpha_1$ and $\beta_1 = \tilde \beta_1$, the amplitudes of the contribution from $\ket{1}_\mathrm{st}\otimes\ket{1}_{\tilde 1}$ cancel.
For the case of $n$ qubits at each node, the above steps are repaeated $n$ times.
The state overlap can then be estimated from the parity of the bitwise AND operation between the measurement outcomes of the storage qubit and the data qubits at the second node.
In principle, these steps may be parallelized at the cost of allowing for additional ancillas.

\subsection{Hong-Ou-Mandel}

\begin{figure}
\centering
\includegraphics[width=0.35\textwidth]{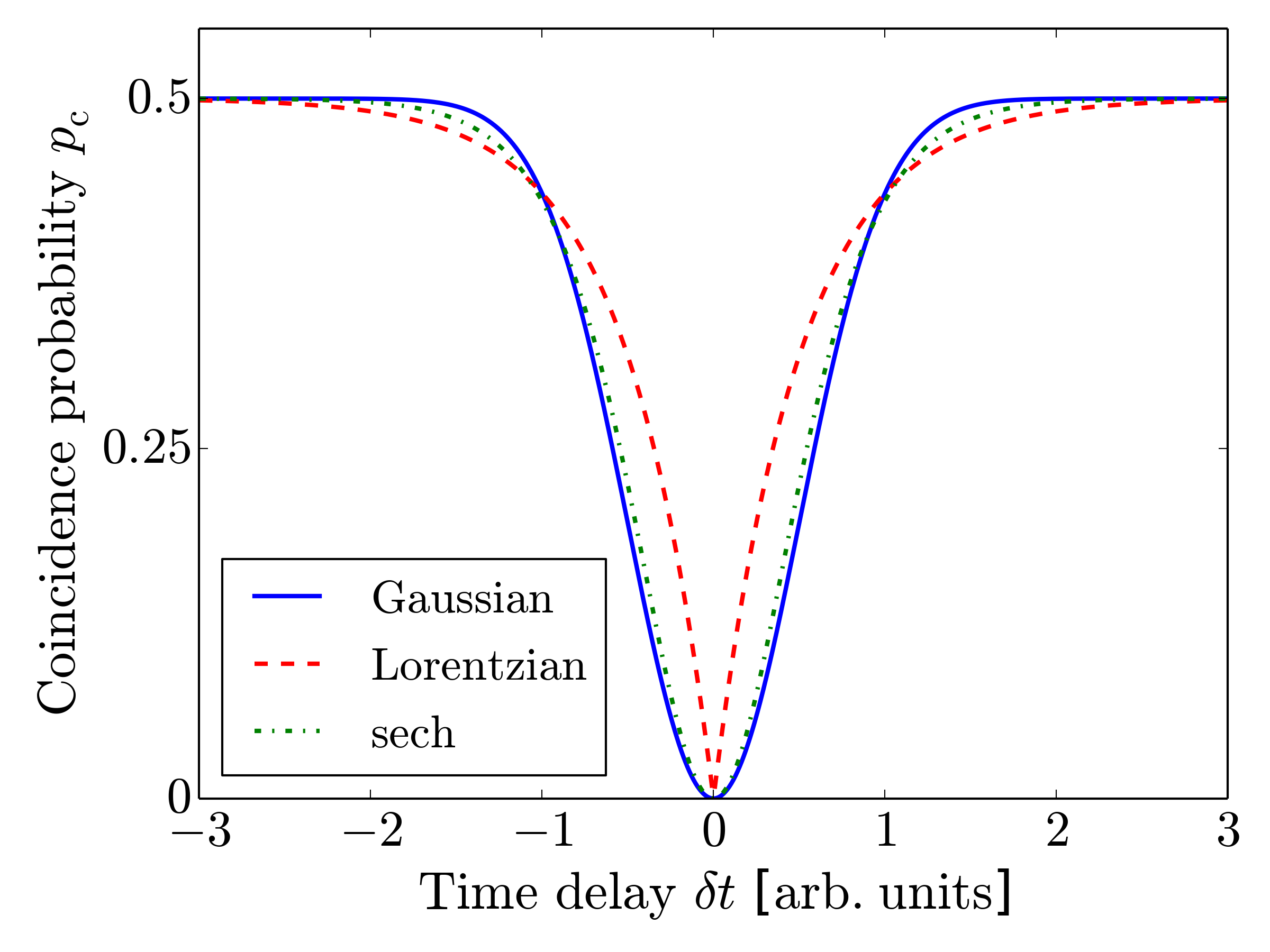}
    \caption{Coincidence probability $p_c$ for different time delays $\delta t$ between arrival times, and otherwise identical photons.
    }
    \label{fig:hom-coincidences}
\end{figure}

The coincidence count rate on the two output detectors of a Hong-Ou-Mandel interference experiment depend on the single-photon temporal mode functions.
In an setup like the one sketched in Fig.~\ref{fig:implementations}, the outcome therefore depends on the qubit-cavity coupling $g(t)$. 
Resulting coincidence count rates can be derived as a function of time delay $\delta t$ between two arriving photons and other physical parameters \cite{Woolley2013}.
For two identical photons, the probability of finding the photons at two different output ports vanishes, $p_\mathrm{c} \rightarrow 0$.
For two distinguishable photons, $p_\mathrm{c} \rightarrow 1/2$.
In an implementation with multiple qubits, this scheme requires an interference experiment between trains of single photons.
Examples for coincidence rates of prototypical mode functions are provided below and shown in Fig.~\ref{fig:hom-coincidences}.

\textit{Gaussian mode function}.
Gaussian mode functions of width $\sigma$ are given by
\begin{equation}
    \psi_{1(2)}(t) = \left ( \frac{1}{\pi\sigma^2} \right )^{1/4} \exp \left [ - \frac{ \left ( t \pm \delta t / 2 \right)^2}{2\sigma^2} \right ]
\end{equation}

The coincidence probability $p_\mathrm{c}$ of finding the photons both at different output detectors can be calculated as
\begin{equation}
    p_\mathrm{c} = \frac{1}{2} \left [ 1 + \exp \left ( - \frac{(\delta t)^2}{2\sigma^2} \right ) \right ]
\end{equation}

\textit{Lorentzian mode function}.
Lorentzian mode functions are characterized by the decay rate $\Gamma$ and are given by
\begin{equation}
    \psi_{1(2)}(t) = \sqrt{2\Gamma} \exp \left [ - \Gamma \left ( t \pm \delta t/2 \right ) \right ] \theta(t \pm \delta t/2 ),
\end{equation}
where $\theta$ denotes the Heaviside step function.
The coincidence probability is then given by
\begin{equation}
    p_\mathrm{c} = \frac{1}{2} \left [ 1 + \exp \left ( - 2\Gamma \delta t \right ) \right ]
\end{equation}

\textit{$\mathrm{sech}$ mode function}.
With the mode function
\begin{equation}
    \psi_{1(2)}(t) = \frac{\sqrt{\Gamma}}{2} \mathrm{sech} \left (\frac{\Gamma \left ( t \pm \delta t/2 \right )}{2} \right ),
\end{equation}
the coincidence probability is found to be given by
\begin{equation}
    p_\mathrm{c} = \frac{\Gamma \delta t}{4} \left ( 1 -  \mathrm{csch} \left ( \frac{\Gamma \delta t}{2} \right ) \right ).
\end{equation}

\section{Estimating confidence intervals \label{app:confidence}}

The quantum state fidelity $\mathcal{F}$ between two states may be estimated by calculating the probability that these states pass one of the overlap tests outlined in the main text, \textit{i.e.}, the distributed SWAP test or a Bell-basis measurement.
We regard the outcomes of either of the tests as a Bernoulli trial in which $X = 0$ denotes \textit{fail} and $X = 1$ denotes \textit{pass}.

\subsection{Hoeffding's inequality}
Large deviation bounds from statistics can be used to determine confidence intervals for the fidelity estimate, such as the Chernoff and Hoeffding bounds.
The latter is well-suited for analytically deriving two-sided concentration bounds.
Let $X_1, ..., X_m$ be independent random variables, where $m$ is the number of repetitions of our experiment, and $X_i \in \{ 0, 1 \}, \forall i \in \{1, ..., m\}$.
With the expectation value $\mathbb{E}[X_i] = \mu$, Hoeffding's inequality takes the form
\begin{equation}
    \Pr \left ( \left | \frac{1}{m} \sum_{j=1}^m X_i - \mu \right | \geq \epsilon \right ) \leq 2 \exp \left ( - 2 m \epsilon^2 \right ).
\end{equation}
The fidelity is related to the success probability by $p_\mathrm{succ} = (1+\mathcal{F})/2$, which leads to Eq.~\eqref{eq:hoeffding}.
Hence, to achieve an $\epsilon$-close approximation of $\mathcal{F}$, with high probability of at least $1 - \alpha$, we require
\begin{equation}
    \alpha \geq 2 \exp\left(-\frac{m\epsilon^2}{2}\right) \Rightarrow m \geq 2 \frac{\ln(2/\alpha)}{\epsilon^2}.
\end{equation}

\subsection{Binomial confidence intervals}

Using Maximum Likelihood Estimation (MLE), we find the most likely value for the fidelity, that is $\hat{\mathcal{F}} = 2p_\mathrm{succ}-1 = 2m_\mathrm{p}/m - 1$, where $m_p$ denotes the the number of successful tests (number of times that the test indicates \textit{pass}).
The Clopper-Pearson interval provides a statistical method for calculating binomial confidence intervals, which we can use to lower- and upper-bound the fidelity at a given confidence level \cite{Beckey2022}.
Passing the test $m_\mathrm{p}$ out of $m$ times means that the success probability will be concentrated around $p_\mathrm{succ} = m_\mathrm{p}/m$.
We obtain a lower and upper bound to this probability, denoted by $p_\uparrow$ and $p_\downarrow$, respectively, at confidence level $1-\alpha$, by solving
\begin{eqnarray}
    \alpha/2 & = &\sum_{j=0}^{m_\mathrm{p}} \binom{m}{j} p_\uparrow(m_\mathrm{p})^j (1-p_\uparrow(m_\mathrm{p}))^{m-j}, \\
    \alpha/2 & = &\sum_{j=m_\mathrm{p}}^{m} \binom{m}{j} p_\downarrow(m_\mathrm{p})^j (1-p_\downarrow(m_\mathrm{p}))^{m-j},
\end{eqnarray}
numerically or express $p_\uparrow(m_\mathrm{p})$ and $p_\downarrow(m_\mathrm{p})$ in terms of quantiles of the $F$ distribution.

From this we can obtain a confidence interval for our fidelity estimate.
The number of times $m_\mathrm{p}$ that two states pass one of the tests out of $m$ total rounds is described by a binomial mass function.
Hence we can use the lower and upper bounds of the $(1-\alpha)$-confidence interval and estimate 
\begin{equation}\label{eq:pearson}
    |\hat{\mathcal{F}} - \mathcal{F}| \leq \epsilon = \sum_{j=0}^m \binom{m}{j} p_\mathrm{succ}^j (1-p_\mathrm{succ})^{m-j} \left [ p_\uparrow(j)-p_\downarrow(j) \right ].
\end{equation}

\begin{figure}[b]
\centering
\includegraphics[width=0.98\textwidth]{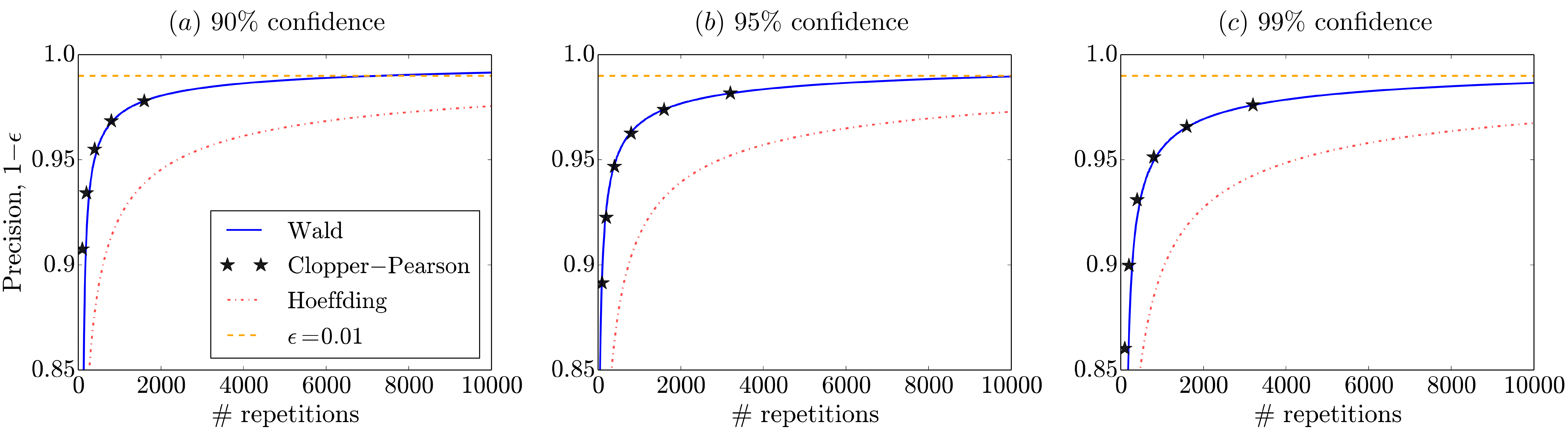}
    \caption{Fidelity $\hat{\mathcal{F}}$ is estimated up to imprecision $\epsilon$ at confidence level $1-\alpha$, where (a) $\alpha = 0.1$, (b), $\alpha = 0.05$, and (c) $\alpha = 0.01$, respectively.
    Hoeffding's bound (\textit{red, dash-dotted line}) is given in the main text in Eq.~\eqref{eq:hoeffding}.
    The Clopper-Pearson confidence interval (\textit{black stars}) is calculated numerically from Eq.~\eqref{eq:pearson}.
    The Wald interval (\textit{blue, solid line}) is derived from the normal approximation to the binomial distribution.
    The dashed orange line shows the threshold for reaching $\epsilon=0.01$ at any given confidence level.
    Other parameters: $\mathcal{F} = (1+\cos(\pi/4))/2$.
    }
    \label{fig:confidence}
\end{figure}

Exemplary results for $\epsilon$ as a function of $m$ are shown in Fig.~\ref{fig:confidence}.
While the Hoeffding bound gives an easily interpretable and analytical result, this numerical approach yields tighter bounds.

\section{Comparison of quantum computations \label{app:comp-state-three-methods}}

Here, we provide some details of the methods discussed in Sec.~\ref{ssec:compare-computations}.
The aim of these methods is to utilize efficient state-comparison schemes to compare remote computations.
We assume two $n$-qubit states with Hilbert space dimensions $d = 2^n$.

\subsection{Method 1: Channel-state duality}
In Fig.~\ref{fig:compare-comps}(a), this method is schematically illustrated for two nodes with a computational (subscript ``$\mathrm{c}$") and an ancillary (``$\mathrm{a}$") register, each.
At each node, both registers are initialized in a maximally entangled state,
\begin{equation}
\ket{\Phi_\sigma^+} = \frac{1}{\sqrt{d}} \sum_{i=0}^{d-1} \ket{i^\sigma_{\mathrm{c}}} \otimes \ket{i^\sigma_\mathrm{a}},
\end{equation}
where $\sigma \in \{ \mathrm{L}, \mathrm{R} \}$ denotes the (\textit{left} and \textit{right}) nodes.

At each node, a computation is performed locally, that is described by a quantum channel $\mathcal{E}_\mathrm{L}$ and $\mathcal{E}_\mathrm{R}$, respectively.
The Choi state at each node then reads
\begin{equation}
\rho_\sigma = \frac{1}{d} \sum_{i,j} \mathcal{E}_\sigma (\ket{i^\sigma_\mathrm{c}}\bra{j^\sigma_\mathrm{c}}) \otimes \ket{i^\sigma_\mathrm{a}}\bra{j^\sigma_\mathrm{a}}.
\end{equation}

For unitary quantum channels, that can be described by the unitary operators $U_\mathrm{L}$ and $U_\mathrm{R}$, respectively, the overlap of the Choi states is simply
\begin{eqnarray}\label{eq:process-fid}
\mathrm{tr} \left ( \rho_\mathrm{L} \rho_\mathrm{R} \right ) & = & \frac{1}{d^2} \mathrm{tr} \left [ \left ( \sum_{i,j} U_\mathrm{L} \ket{i^\mathrm{L}_\mathrm{c}}\bra{j^\mathrm{L}_\mathrm{c}} U_\mathrm{L}^\dagger \otimes \ket{i^\mathrm{L}_\mathrm{a}}\bra{j^\mathrm{L}_\mathrm{a}} \right )
\left ( \sum_{k,l} U_\mathrm{R} \ket{k^\mathrm{R}_\mathrm{c}}\bra{l^\mathrm{R}_\mathrm{c}} U_\mathrm{R}^\dagger \otimes \ket{k^\mathrm{R}_\mathrm{a}}\bra{l^\mathrm{R}_\mathrm{a}} \right )\right ] \\
& = & \frac{1}{d^2} \mathrm{tr} \left [ \sum_{i,j,k,l}
\left ( U_\mathrm{L} \ket{i^\mathrm{L}_\mathrm{c}}\bra{j^\mathrm{L}_\mathrm{c}} U_\mathrm{L}^\dagger  U_\mathrm{R} \ket{k^\mathrm{R}_\mathrm{c}}\bra{l^\mathrm{R}_\mathrm{c}} U_\mathrm{R}^\dagger  \right ) \otimes
\left ( \ket{i^\mathrm{L}_\mathrm{a}}\braket{j^\mathrm{L}_\mathrm{a}|k^\mathrm{R}_\mathrm{a}}\ket{l^\mathrm{R}_\mathrm{a}} \right )
\right ] \nonumber \\
& = & \frac{1}{d^2} \left | \sum_i \braket{i|U_\mathrm{R}^\dagger U_\mathrm{L}|i} \right |^2 = \frac{1}{d^2} \left | \mathrm{tr}  ( U_\mathrm{L}^\dagger U_\mathrm{R} ) \right |^2. \nonumber
\end{eqnarray}
This overlap can be estimated using one of the protocols for distributed overlap estimation.

\subsection{Method 2: Random initial states \label{app:random-initial-states}}

\subsubsection{Random unitaries from unitary 2-design}

The final expression in Eq.~\eqref{eq:process-fid} is also refered to as the process fidelity:
\begin{equation}
\mathcal{F}_\mathrm{p}(U_\mathrm{L}, U_\mathrm{R}) = \frac{\left|\mathrm{tr}\left(\uld\ur\right)\right|^2}{d^2}.
\end{equation}
On the other hand, the average fidelity $\mathcal{F}_\mathrm{av}$ of a quantum channel $\mathcal{E}$ and a unitary quantum gate $U$ can be defined as
\begin{equation}
    \mathcal{F}_\mathrm{av}(\mathcal{E}, U) = \int \mathrm{d} \psi \ \langle\psi| U^\dagger \mathcal{E}(|{\psi}\rangle\langle{\psi}|) U |\psi\rangle,
\end{equation}
which quantifies how well $U$ is approximated by $\mathcal{E}$.
Here, $\mathrm{d}\psi$ is the uniform Haar measure on state space.
For a unitary channel, this leads to an expression for the average fidelity between unitaries,
\begin{equation}\label{eq:av-fid}
    \mathcal{F}_\mathrm{av}(\ul, \ur) = \int \mathrm{d} \psi \ |\langle\psi| \uld \ur |\psi\rangle |^2,
\end{equation}
which compares the ability of $\ul$ and $\ur$ on average to make states orthogonal \cite{Acin2001}.
The expressions in Eq.~\eqref{eq:process-fid} and \eqref{eq:av-fid} are related by \cite{nielsen01}
\begin{equation}\label{eq:fid-relation}
\mathcal{F}_\mathrm{p}(\ul, \ur) = \frac{(d+1) \mathcal{F}_\mathrm{av}(\ul, \ur) - 1}{d}.
\end{equation}

The random initial states in \eqref{eq:av-fid} may be generated by applying Haar-random unitaries to the fiducial state $\ket{\psi_0}$:
\begin{eqnarray}\label{eq:av-design}
\mathcal{F}_\mathrm{av}(\ul,\ur) & = & \int_\mathcal{U} \mathrm{d}\mu(U) |\langle \psi_0 | U^\dagger \uld \ur U | \psi_0 \rangle|^2 \nonumber \\
& = & \frac{1}{K} \sum_{k=1}^K |\langle \psi_0| U_k^\dagger \uld \ur U_k | \psi_0 \rangle|^2,
\end{eqnarray}
where in the last step we have used the fact that $\mathrm{d}U$ has support only on a finite set of unitaries.
The unitaries $\{ U_k \}_{k \in \{1, ..., K \} }$ form a unitary $2$-design.

The average fidelity can be estimated by sampling unitaries uniformly at random, and measuring the overlap in \eqref{eq:av-design} via one of the abovementioned protocols.
The process fidelity can be estimated from the averaged fidelity and by means of \eqref{eq:fid-relation}, and the confidence intervals can be estimated using standard tools from statistical inference.

\subsubsection{Computational basis sampling}
Here we examine the similarity of two unitary operations $\ul$ and $\ur$ by comparing the states $\ul |p\rangle$ and $\ur |p\rangle$ for different initial states $|p\rangle$.
For the initial states, we use the computational basis.
The measure
\begin{eqnarray}\label{eq:fid-sq}
    \mathcal{F}_\mathrm{sq} = \frac{1}{d} \sum_{p\in\{0,1\}^n} |\langle p | U_L^\dagger U_R | p \rangle |^2
\end{eqnarray}
equals one if $\ul = \ur =: U$, and $\mathcal{F}_\mathrm{sq} < 1$ otherwise.
By sampling $m_b$ initial states uniformly at random from the computational basis, we define
\begin{equation}\label{eq:fsq-approx}
\hat{\mathcal{F}}_\mathrm{sq} := \mathcal{F}_\mathrm{sq}^{(m_b)} = \frac{1}{m_b} \sum_{p \in \mathcal{S}} |\langle p | \uld \ur | p \rangle|^2,
\end{equation}
where $\mathcal{S}$ denotes the set of $m_b$ basis states that are contained in our samples.
Each term in the sum requires sufficiently many samples to obtain a good approximation of $|\langle p | \uld \ur | p \rangle |^2$.
Using large-deviation bounds, cf.~App.~\ref{app:confidence}, to achieve an $\epsilon$-close approximation of each term we need $m_s = O(1/\epsilon^2)$ samples.
If each term in Eq.~\eqref{eq:fsq-approx} is estimated up to some error $\epsilon$, the propagated uncertainty of the sum is $\epsilon/\sqrt{m_b}$.
A $\epsilon_m$-close approximation of $\mathcal{F}_\mathrm{sq}^{(m_b)}$ thus requires $m_b \cdot m_s = O(1/\epsilon_m^2)$ repetitions of the state-comparison experiment.
By means of Hoeffding's inequality, it suffices to sample only a few basis states, independent of system size.

\begin{figure}[b]
\centering
\includegraphics[width=1.0\textwidth]{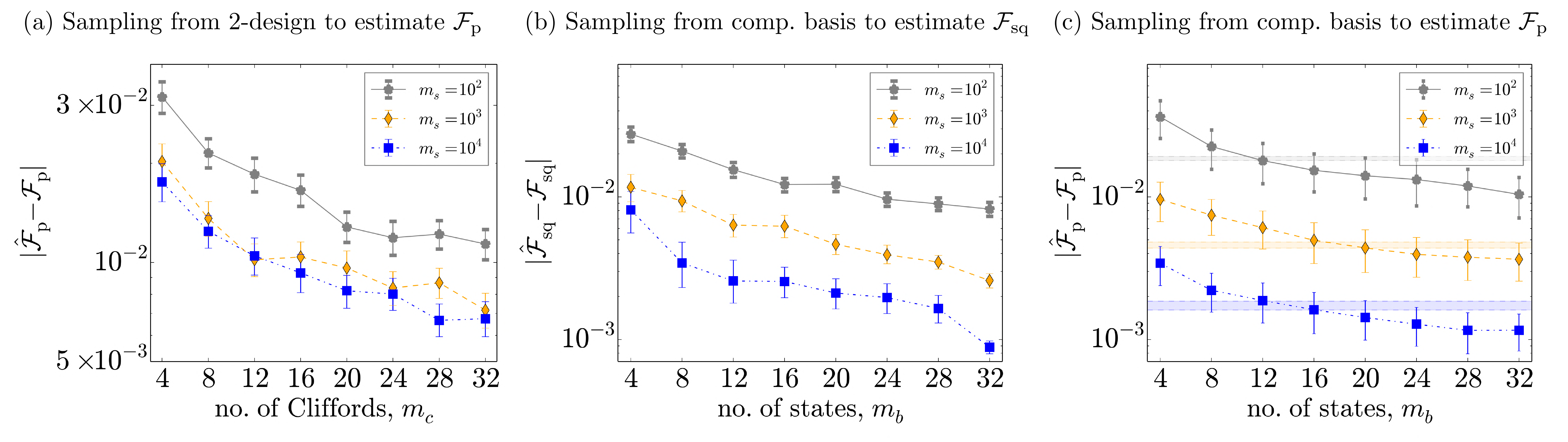}
    \caption{Numerical results obtained with the methods in App.~\ref{app:random-initial-states}.
    }
    \label{fig:app-random-numerics}
\end{figure}

In the numerical example from the main text, we implement $\ul = U$ and $\ur = M U$ where $M$ is a product of random single-qubit rotations.
Eq.~\eqref{eq:fsq-approx} becomes a good approximation for sufficiently many samples.
Numerical results for a $5$-qubit example are shown in Fig.~\ref{fig:app-random-numerics}(b).

\subsubsection{Hadamard tests}

\begin{figure*}
\begin{picture}(3,3)
\put(0,45.0) {{$(a)$}}
\put(130.0,45.0) {{$(b)$}}
\put(300.0,45.0) {{$(c)$}}
\end{picture}
 \resizebox{0.25\textwidth}{!}{%
\begin{quantikz}
\lstick[wires=1]{$\ket{+}$} & \ctrl{1} & \gate{H} & \meter{} & \cw \rstick{$a$} \\
\lstick[wires=1]{$\ket{\psi}$} & \gate{U}\qwbundle[alternate]{} & \qwbundle[alternate]{} & \qwbundle[alternate]{} & \qwbundle[alternate]{} \\\begin{picture}(3,3)
\put(0,45.0) {{$(a)$}}
\put(130.0,45.0) {{$(b)$}}
\put(300.0,45.0) {{$(c)$}}
\end{picture}
\end{quantikz}
}
\hspace{-0.1cm}
 \resizebox{0.3\textwidth}{!}{%
\begin{quantikz}
\lstick[wires=1]{$\ket{+}$} & \octrl{1} & \ctrl{1} & \gate{H} & \meter{} & \cw \rstick{$b$} \\
\lstick[wires=3]{$\ket{p}$} & \gate[3, nwires=2]{U_\mathrm{L}} & \gate[3, nwires=2]{U_\mathrm{R}} & \qw & \qw & \qw \\
\ \ \vdots\ \\
& \qw & \qw & \qw & \qw & \qw \\
\end{quantikz}
}
\hspace{-0.1cm}
 \resizebox{0.35\textwidth}{!}{%
\begin{quantikz}
\lstick[wires=1]{$\ket{+}$} & \octrl{1} & \ctrl{2} & \gate{H} & \meter{} & \cw \rstick{$c$} \\
\lstick[wires=2]{$\ket{\Phi^+}^{\otimes n}$}  & \gate{U_\mathrm{L}} \qwbundle[alternate]{} & \qwbundle[alternate]{} & \qwbundle[alternate]{} \\
 & \qwbundle[alternate]{} & \gate{U_\mathrm{R}} \qwbundle[alternate]{} & \qwbundle[alternate]{} \\
\end{quantikz}
}
\caption{Circuit representation of (a) standard and (b-c) modified Hadamard tests.
}
\label{fig:hadamard_tests}
\end{figure*}
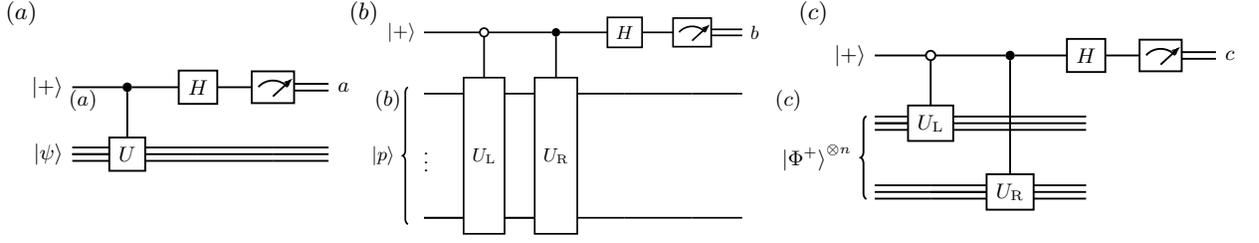

The normalized trace of a $d \times d$ unitary $U$ may be estimated with a standard Hadamard test, cf.~Fig.~\ref{fig:hadamard_tests}(a), using an ancilla qubit initialized in the $\ket{+}$ state, and the $n$-qubit computational register in the maximally mixed state.
After applying the controlled-$U$ operator to $(|+\rangle\langle +|/2) \otimes (\mathrm{Id}_d/d)$, and subsequently applying a Hadamard gate to the ancilla qubit, the probability of measuring the ancilla qubit in $0$ becomes $[1 + \mathrm{Re} ~ \mathrm{tr}(U)/d ]/2$.
It can be estimated with accuracy $\epsilon$ and confidence level $(1-\alpha)$ using $O(\mathrm{ln}(2/\alpha)/\epsilon^2)$ repetitions.
The imaginary part of $\mathrm{tr}(U)/d$ can be estimated similarly by initializing the ancilla in $(\ket{0} - \mathrm{i} \ket{1})/\sqrt{2}$.

Choosing the initial state of the computational register uniformly at random from the computational basis $\{ \ket{0, 0, ..., 0}, \ket{1, 0, ..., 0}, ..., \ket{1, ..., 1} \}$ is equivalent to inputting the maximally mixed state to the computational register \cite {Datta2005,Shor2007}.
The normalized trace may thus be estimated by sampling from the computational basis set.

\textit{Modified Hadamard test}.\textemdash
Before the measurement, the modified Hadamard test from Fig.~\ref{fig:hadamard_tests}(b) yields the state
\begin{equation}
\ket{\Psi}_\mathrm{tot} = \frac{1}{2} \ket{0}_\mathrm{anc} \otimes ( \ul \ket{p} + \ur \ket{p}) + \frac{1}{2} \ket{1}_\mathrm{anc} \otimes ( \ul \ket{p} - \ur \ket{p} ),
\end{equation}
such that the measurement output is $0$ with probability
\begin{equation}
    p_0^{(r)} = \frac{1 + \mathrm{Re}~\langle p | \uld \ur | p \rangle}{2}.
\end{equation}
Similarly, initializing the ancilla qubit in $\ket{-i} = (\ket{0} - i \ket{1})/\sqrt{2}$ with an additional $S^\dagger$ gate, the measurement result will be $0$ with probability $p_0^{(i)} = (1 + \mathrm{Im}~\langle p | \uld \ur | p \rangle )/2$.
Together, this allows us to estimate
\begin{equation}
\langle p | \uld \ur | p \rangle = 2(p_0^{(r)} + p_0^{(i)} \cdot \mathrm{i}) - ( 1 + \mathrm{i}).
\end{equation}

As before, here we consider an example where $\ul = U$ and $\ur = M U$, where $M$ is a product or random single-qubit rotations.
The normalized trace can be estimated efficiently up to some error $\epsilon$ by sampling $O(1/\epsilon^2)$ initial basis states $\ket{p_k}$ uniformly at random from the computational basis, \textit{i.e.}, $\mathrm{tr}(\uld \ur)/d \approx (1/m_\mathrm{b}) \sum_{k=1}^{m_\mathrm{b}} \langle p_k | \uld \ur | p_k \rangle$.
Numerical results for a $5$-qubit example are shown in Fig.~\ref{fig:app-random-numerics}(c), for $m_s = 10^2, 10^3, 10^4$ repetitions of the Hadamard test per basis state and $m_b \in \{4, 8, ..., 32 \}$ sampled basis states.

To realize the test from Fig.~\ref{fig:hadamard_tests}(b) between two $n$-qubit unitaries in a distributed setting with remote devices, $(n+1)$ qubit transmissions are required, because the ancilla qubit and an entire register need to be transmitted.

\textit{Using maximally entangled registers.}\textemdash
Alternatively, if the two registers are in a maximally entangled state, the Hadamard test may be performed using the circuit in Fig.~\ref{fig:hadamard_tests}(c) and transmitting a single qubit.
Before the measurement, the Hadamard test from Fig.~\ref{fig:hadamard_tests}(c) yields the state
\begin{equation}
    \ket{\Psi}_\mathrm{tot} = \frac{1}{2\sqrt{d}} \ket{0}_\mathrm{anc} \left ( \sum_j \ul \ket{j}\otimes \ket{j} + \ket{j} \otimes \ur \ket{j} \right )
    + \frac{1}{2\sqrt{d}} \ket{1}_\mathrm{anc} \left ( \sum_j \ul \ket{j}\otimes \ket{j} - \ket{j} \otimes \ur \ket{j} \right ),
\end{equation}
such that the measurement output is $0$ with probability
\begin{equation}
    p_0^{(r)} = \frac{1}{2} + \frac{1}{2d} \mathrm{Re} \sum_j \langle j | \ul^* \ur | j \rangle.
\end{equation}
Similarly, initializing the ancilla qubit in $\ket{-i} = (\ket{0} - i \ket{1})/\sqrt{2}$ with an additional $S^\dagger$ gate, the measurement result will be $0$ with probability $p_0^{(i)} = (1 + \mathrm{Im}~\sum_i \langle i | \ul^* \ur | i \rangle/d )/2$.
Together, this allows us to estimate
\begin{equation}
    \frac{1}{d} \sum_{j=0}^{d-1} \langle j | \ul^* \ur | j \rangle = 2 (p_0^{(r)} + p_0^{(i)} \cdot \mathrm{i}) - (1 + \mathrm{i}).
\end{equation}
If Alice is supposed to implement a unitary $\ul = U^T$ on her device, and Bob implements $\ur = U$, we can detect deviations form a perfect implementation, since
\begin{equation}
    4 | (p_0^{(r)} - 1) |^2 + 4 | p_0^{(i)} - 1 |^2 = \frac{|\mathrm{tr}(\ul^* \ur)|^2}{d^2} = \frac{|\mathrm{tr}(U^\dagger U)|^2}{d^2} = 1,
\end{equation}
if and only if $\ul = \ur^T$.

Numerical results of this test for the same single-qubit rotation example described above are included in Fig.~\ref{fig:app-random-numerics}(c).

\subsection{Method 3: Entanglement tests}
The maximally entangled state $\ket{\Phi^+} = \sum_j |j, j\rangle / \sqrt{d}$ remains invariant under transformations of the form $U \otimes U^*$, \textit{i.e.},
\begin{equation}
    \left ( U \otimes U^* \right ) \ket{\Phi^+} = \left ( \mathbb{1} \otimes U^\mathrm{T} U^* \right ) \ket{\Phi^+} = \left ( \mathbb{1} \otimes (U^*)^\dagger U^* \right ) \ket{\Phi^+} = \ket{\Phi^+}.
\end{equation}
From this it follows that if Alice applies $U$, then Bob can apply $U^*$ and the state remains unchanged.
More generally, the istropic state, \textit{i.e.}, the mixture of completely depolarized noise and the maximally entangled state,
\begin{equation}
    \rho_\alpha^{(d)} = \alpha |\Phi^+\rangle\langle\Phi^+| + \frac{1-\alpha}{d^2} \mathbb{1}\otimes\mathbb{1},
\end{equation}
with $-1/(d^2-1) \leq \alpha \leq 1$ remains invariant under $U \otimes U^*$:
\begin{eqnarray}
    \left ( U \otimes U^* \right ) \rho_\alpha^{(d)} \left ( U \otimes U^* \right )^\dagger = \rho_\alpha^{(d)}.
\end{eqnarray}

If $U$ and $U^*$ are not perfectly implemented, the final state is not the maximally entangled state.
This will be detectable in a CHSH experiment, as demonstrated in the example given below.

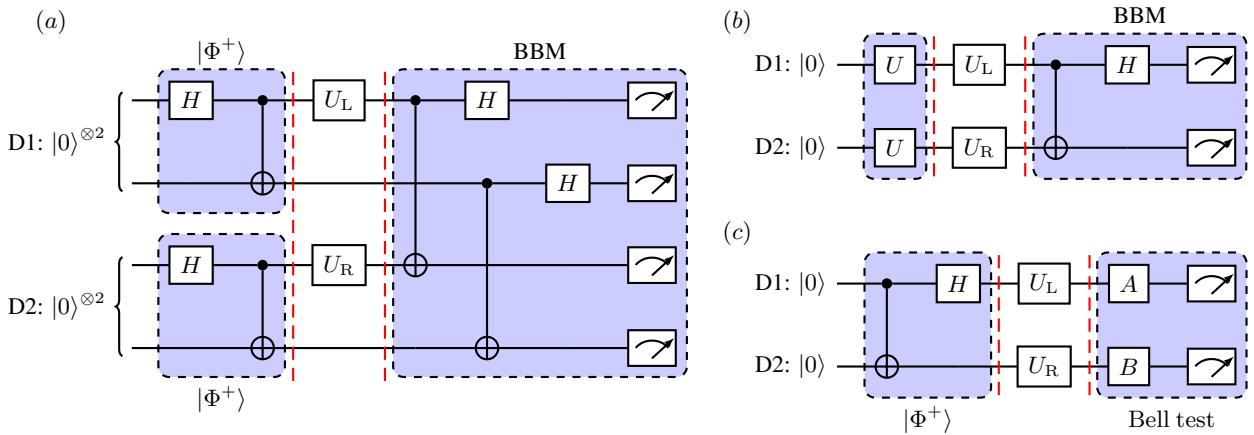
\begin{figure*}[b]
\begin{minipage}{.45\textwidth}
\begin{picture}(3,3)
\put(-100,0.0) {{$(a)$}}
\put(160.0,0.0) {{$(b)$}}
\put(160.0,-80.0) {{$(c)$}}
\end{picture}
\begin{quantikz}
\lstick[wires=2]{D1: $\ket{0}^{\otimes 2}$} & \gate{H} \gategroup[wires=2,steps=2,style={dashed,rounded corners,fill=blue!20,inner sep=1pt},background]{$\ket{\Phi^+}$} & \ctrl{1} \slice{} & \gate{U_\mathrm{L}} \slice{} &
\ctrl{2} \gategroup[wires=4,steps=4,style={dashed,rounded corners,fill=blue!20,inner sep=1pt},background]{BBM} & \gate{H} & \qw & \meter{} & \\
 & \qw & \targ{} & \qw & \qw & \ctrl{2} & \gate{H} & \meter{} \\
\lstick[wires=2]{D2: $\ket{0}^{\otimes 2}$} & \gate{H} \gategroup[wires=2,steps=2,style={dashed,rounded corners,fill=blue!20,inner sep=1pt},background, label style={label position=below, yshift = -0.5cm}]{$\ket{\Phi^+}$} & \ctrl{1} & \gate{U_\mathrm{R}} & \targ{} & \qw & \qw & \meter{} \\
 & \qw & \targ{} & \qw & \qw & \targ{} & \qw & \meter{}
\end{quantikz}
\end{minipage}
\hspace{1cm}
\begin{minipage}{.45\textwidth}
\begin{quantikz}
\lstick[wires=1]{D1: $\ket{0}$} & \gate{U}\gategroup[wires=2,steps=1,style={dashed,rounded corners,fill=blue!20,inner sep=1pt},background, label style={label position=above, yshift = -0.5cm}]{} \slice{} & \gate{U_\mathrm{L}} \slice{} & \ctrl{1} \gategroup[wires=2,steps=3,style={dashed,rounded corners,fill=blue!20,inner sep=1pt},background]{BBM} & \gate{H} & \meter{} \\
\lstick[wires=1]{D2: $\ket{0}$} & \gate{U} & \gate{U_\mathrm{R}} & \targ{} & \qw & \meter{}
\end{quantikz}
\vspace{0.6cm}
\begin{quantikz}
\lstick[wires=1]{D1: $\ket{0}$} & \ctrl{1} \gategroup[wires=2,steps=2,style={dashed,rounded corners,fill=blue!20,inner sep=1pt},background, label style={label position=below, yshift = -0.5cm}]{$\ket{\Phi^+}$} & \gate{H}\slice{} & \gate{U_\mathrm{L}}\slice{} & \gate{A} \gategroup[wires=2,steps=2,style={dashed,rounded corners,fill=blue!20,inner sep=1pt},background, label style={label position=below, yshift = -0.5cm}]{$\mathrm{Bell}$ $\mathrm{test}$} & \meter{} \\
\lstick[wires=1]{D2: $\ket{0}$} & \targ{} & \qw & \gate{U_\mathrm{R}} & \gate{B} & \meter{}
\end{quantikz}
\end{minipage}
\caption{Circuits for
(\textit{a}) comparing Choi states,
(\textit{b}) random initial states $U\ket{0}$ for $U$ sampled from a unitary 2-design,
(\textit{c}) entanglement tests.
In (a) and (b), the fidelity is estimated using Bell-basis measurements (BBM).
In (c), the fidelity is estimated by extracting the expectation value of the CHSH operator $S$ defined in Eq.~\eqref{eq:s-operator}.
}
\label{fig:comp-example-circuits}
\end{figure*}

\subsection{Example}
Here, we provide a simple example, which we implement on a IBM device.
In analogy to the previous example provided in Fig.~\ref{fig:fidelity}, we compare the implemented versions of the single-qubit unitaries $\ul = \mathrm{H}$ and $\ur = \mathrm{P}(\phi) \mathrm{H}$, where $\mathrm{P}(\phi) = \left (\begin{matrix}
    1 & 0 \\
    0 & e^{i\phi}
\end{matrix}\right )$.
Then we compare all experimental results to the ideal value $\mathcal{F}_\mathrm{p} = (1+\cos\phi)/2$.

The circuits of all performed experiments are provided in Fig.~\ref{fig:comp-example-circuits}.
Fig.~\ref{fig:comp-example-circuits}(a) shows the circuit for comparing two two-qubit Choi states and (b) shows the circuit for estimating the fidelity via an averaged fidelity for different initial states.
Sampling of random unitaries becomes useful at larger system sizes, and in this example we can run all circuits necessary to obtain all fidelities in \eqref{eq:av-design}.
Fig.~\ref{fig:comp-example-circuits}(c) shows the circuit for performing a Bell test on the state $(\ul \otimes \ur) |\Phi^+\rangle$.
We define the CHSH operator as
\begin{equation}\label{eq:s-operator}
    S = \left ( A_0 \otimes B_0 \right ) - \left ( A_0 \otimes B_1 \right ) + \left ( A_1 \otimes B_0 \right ) + \left ( A_1 \otimes B_1 \right ),
\end{equation}
with $A_0 = (\mathrm{Z}-\mathrm{X})/\sqrt{2}$, $A_1 = (\mathrm{X}+\mathrm{Z})/\sqrt{2}$,
$B_0 = \mathrm{Z}$, $B_1 = \mathrm{X}$.
It holds that $\langle \Phi^+ | S | \Phi^+ \rangle = 2 \sqrt{2}$.
After applying $\ul$ and $\ur$, the expectation value is $\langle \Phi^+ | H^\dagger P^\dagger S H | \Phi^+ \rangle = \sqrt{2} ( 1 + \cos \phi)$.
The results of the experiments after post-processing are shown in Fig.~\ref{fig:compare-comps}(d) in the main text.

\end{document}